\documentclass[preprint,noshowkeys,amsmath,amssymb,superscriptaddress,nofootinbib]{revtex4-1}

\usepackage{bm}
\usepackage{slashed}

\usepackage[pdftex]{graphicx,color}
\usepackage{epstopdf}
\usepackage[bookmarks=true,bookmarksnumbered=true,bookmarkstype=toc, colorlinks=true, 
citecolor=purple, linkcolor=purple ,urlcolor=purple]{hyperref} 
\usepackage{units}

\newcommand{\lsim}{\raise0.3ex\hbox{$\;<$\kern-0.75em\raise-1.1ex\hbox{$\sim\;$}}}
\newcommand{\gsim}{\raise0.3ex\hbox{$\;>$\kern-0.75em\raise-1.1ex\hbox{$\sim\;$}}}
\usepackage{mathtools}
\newcommand{\Slash}[1]{{\ooalign{\hfil/\hfil\crcr\(#1\)}}}

\renewcommand{\eqref}[1]{Eq.~(\ref{#1})}

\usepackage{color}

\definecolor{green}{cmyk}{1,0,1,0}

\definecolor{pink}{cmyk}{0,0.5,0,0}

\definecolor{pastelpink}{cmyk}{0,0.25,0,0}

\definecolor{softpink}{cmyk}{0,0.125,0,0}

\definecolor{purple}{cmyk}{0.5,1.0,0.1,0}

\definecolor{violet}{cmyk}{0.75,1,0.25,0}

\usepackage{ulem}  

\usepackage[utf8]{inputenc}

\begin{document}

\preprint{UME-PP-020}
\preprint{KYUSHU-HET-232}

\title{Inelastic Dark Matter from Dark Higgs Boson Decays at FASER}

\author{Jinmian Li}
\email{jmli@scu.edu.cn}
\affiliation{College of Physics, Sichuan University, Chengdu 610065, China}

\author{Takaaki Nomura}
\email{nomura@scu.edu.cn}
\affiliation{College of Physics, Sichuan University, Chengdu 610065, China}

\author{Takashi Shimomura}
\email{shimomura@cc.miyazaki-u.ac.jp}
\affiliation{Faculty of Education, University of Miyazaki, 
1-1 Gakuen-Kibanadai-Nishi, Miyazaki 889-2192, Japan}
\affiliation{Department of Physics, Kyushu University, 744 Motooka, Nishi-ku, Fukuoka, 819-0395, Japan}

\begin{abstract}
We consider inelastic dark matter scenarios with dark photon mediator and a dark Higgs boson. 
The dark Higgs boson spontaneously breaks the gauge symmetry associated with the dark photon, and gives 
the mass to the dark photon and the mass difference to dark particles. 
Such a dark Higgs boson can decay into the dark particles and hence can be another source of the dark particles 
at collider experiments. We analyze the sensitivity to decays of the excited state into the dark matter and 
charged particles at the FASER 2 experiment in fermion and scalar inelastic dark matter scenarios. We consider 
two mass spectra as illustrating examples in which the excited state can be produced only through the decay 
of dark Higgs boson. 
We show that unprobed parameter region can be explored in fermion dark matter scenario for the illustrating 
mass spectra.
\end{abstract}

\date{\today}

\maketitle

\section{Introduction}
Dark matter (DM) is one of the fundamental questions in particle physics and cosmology. 
The existence of DM has been confirmed through gravitational phenomena in astrophysical observations.
However, its nature still remains a mystery, except for the relic abundance at the Universe today  \cite{Planck:2018vyg}. 
Many hypothetical particles have been proposed for DM candidates and those have been 
searched by direct and indirect experiments over the past decade (for a review, e.g. \cite{Battaglieri:2017aum}).
Null results from these experiments have set bounds on their cross sections with ordinary matters 
over wide range of mass. Future experiments will explore the rest of parameter space in DM landscape.

Among the DM candidates, inelastic DM (iDM) is a compelling candidate for sub-GeV thermal 
dark matter \cite{Tucker-Smith:2001myb,Tucker-Smith:2004mxa}, which was originally motivated by 
the annual modulation reported from the DAMA/LIBRA experiment \cite{DAMA:2008jlt,Bernabei:2013xsa,DAMA:2010gpn,universe4110116}.\footnote{Recently, 
the iDM is also applied to 
XENON1t excess~\cite{Harigaya:2020ckz,Baek:2020owl,Kim:2020aua,Borah:2020jzi,Baek:2021yos,Dutta:2021wbn}.} 
In the iDM scenario, there are two dark particles with different masses i.e. the lighter dark matter state and 
heavier excited state. Elastic interactions of both state to mediator particle are assumed to 
be absent or much suppressed, and inelastic one dominantly occurs in scatterings. 
Therefore the DM state inelastically scatters off the Standard Model (SM) particles through 
the exchange of the mediator and is converted to the excited state, or vice versa. 
Due to this property, for a large mass splitting between two states, constraints from direct detection 
experiment and residual DM annihilations from the Cosmic Microwave Background can be evaded. 
Furthermore, when the mediator particle has a sub-GeV mass, light thermal dark matter below Lee-Weinberg bound 
\cite{Lee:1977ua} can be realized.

In a simple and widely studied setup, the dark matter and excited state only couple 
to a massive dark photon which is the gauge boson of a new U(1) gauge symmetry. 
The dark photon connects the dark and SM particles through the mixing with the hypercharge gauge boson or the electromagnetic photon \cite{Holdom:1985ag,Fayet:1990wx}. 
In the literature, the mass of the dark photon is generated in the Higgs mechanism with a dark Higgs boson. As 
a consequence of the spontaneous symmetry breaking, the dark Higgs boson has an interaction which 
leads decay of the dark Higgs boson into two dark photons if kinematically allowed. On the other hand, 
depending on the gauge charges, the symmetry can allow such a dark Higgs boson to couple to the dark particles.  
In this case, the mass difference between the dark matter and excited states can be generated 
by the symmetry breaking. Then the inelastic interactions for the dark particles with the dark photon 
are also generated. 
Thus, the origins of the dark photon mass and inelastic interaction can be simultaneously explained 
in these models.

In fixed target and beam dump experiments, the dark photon can be produced in meson decays and bremsstrahlung of 
charged particles (see a review \cite{Bauer:2018onh} for references). 
When the dark photon is enough heavy, it decays into the pair of the dark matter and/or the excited state. 
The produced excited state then decays into the DM and the SM charged particles. 
Such decays have been searched as visible decays in \cite{PhysRevD.38.3375,Darme:2018jmx,LSND:2001akn}. 
In the case that the excited state is 
too long-lived to decay in detector,  it has been searched as missing energy in 
\cite{BaBar:2017tiz,NA64:2017vtt,Banerjee:2019pds} 
or scattering off electrons in detector \cite{PhysRevD.38.3375,Batell:2014mga}. 
The results from these experiments gave the constraints on the mass and coupling constants \cite{Jodlowski:2019ycu}. 
Then, in the present allowed parameter space, 
the dark photon and the excited states become long-lived, which can be searched in various on-going 
and planned experiments. 
Projection of sensitivities to such long-lived particles have been studied for 
Belle-II \cite{Izaguirre:2015zva,Duerr:2019dmv, Duerr:2020muu,Kang:2021oes,Dreyer:2021aqd}, 
FASER \cite{Berlin:2018jbm, Jodlowski:2019ycu,Batell:2021blf}, 
MATHUSLA \cite{Berlin:2018jbm,Jodlowski:2019ycu,Guo:2021vpb} 
and short baseline neutrino program \cite{Batell:2021ooj}, respectively. 
In these studies, the main production process 
of the excited state is the decay of the dark photon. 
However, when the dark particles are coupled to the dark Higgs boson, 
these also can be produced from the decay of the dark Higgs boson which can be produced 
from $B$ and $K$ meson decays. 
This production opens possibilities to probe mass spectra in which the excited state is too heavy 
to be produced from the dark photon decay. Such mass spectra has not been analyzed yet, simply due to 
the small number of the excited state through off-shell dark photon. 
In the LHC experiment, a huge number of the mesons are produced due to high 
center-of-mass energy and luminosity. Hence a large number of the dark Higgs boson decays are expected, and its subsequent decay products can be probed by the FASER detector \cite{Feng:2017uoz,FASER:2018bac}. 
Since the production from the dark Higgs decay is the consequence of the dark particle mass generation, 
it is important to analyze the sensitivity to that production process.

In this paper, we study the sensitivity of the search for iDM with sub-GeV mass at the FASER experiment. 
To focus on the importance of the dark Higgs decays, 
we consider two illustrating mass spectra for fermion and scalar iDM in which the dark particles are only produced from the dark Higgs, and analyze the sensitivity to the visible decays of the excited state at the FASER experiment.
This paper is organized as follows. In section \ref{sec:model}, we introduce our models of fermion and scalar 
inelastic dark matter, and give the masses and relevant interactions, respectively. In section \ref{sec:analysis}, we show the numerical results of the sensitivity. We conclude our study in section \ref{sec:conclusion}.

\section{models} \label{sec:model}

We consider two iDM models for a fermion and scalar DM, respectively, with a gauge boson of local $U(1)_X$ symmetry. 
We denote each dark matter candidate as
\begin{itemize}
\item Dirac fermion $\chi$,
\item Complex scalar $S = \frac{1}{\sqrt{2}} (s + i a)$,
\end{itemize}
where both $\chi$ and $S$ have $U(1)_X$ charge $+\frac{1}{2}$ but are SM singlets. All SM particles are assumed to be 
neutral under the $U(1)_X$ symmetry.
To break the $U(1)_X$ symmetry, we also introduce a SM singlet scalar field $\varphi$ with $U(1)_X$ charge $+1$. 
With this charge assignment, the scalar field can form Yukawa or cubic interaction terms with $\chi$ and $S$.
After the spontaneous symmetry breaking, the dark photon acquires a mass, and the models have a remnant $Z_2$ symmetry 
where $\chi/S$ is odd and the other particles are even under it. Furthermore, the dark matter candidates split 
into two mass eigenstates due to the interaction terms. In the following, we give the Lagrangian of the models, the mass 
and mixing, interaction terms of the dark matter candidates. 

\subsection{Lagrangian} \label{subsec:lagrangian}
The Lagrangian of our models is given by
\begin{align}
\mathcal{L}  = \mathcal{L}_{SM} + \mathcal{L}^{\chi(S)}_{DM} - \frac{1}{4} X^{\mu \nu} X_{\mu \nu} - \frac{\epsilon}{2} B_{\mu \nu} X^{\mu \nu}  + (D^\mu \varphi)^\ast (D_\mu \varphi)  - V, 
\end{align}
where $\mathcal{L}_{SM}$ is the SM Lagrangian without the Higgs potential and 
$\mathcal{L}^{\chi(S)}_{DM}$ is the Lagrangian for our fermion(scalar) iDM scenario shown below.
The gauge fields of $U(1)_X$ and $U(1)_Y$ are denoted by $X$ and $B$, and the same symbols are used 
for their field strengths.
The forth term represents the gauge kinetic mixing with a constant parameter $\epsilon$. 
The covariant derivative is given by 
\begin{equation}
D_\mu = \partial_\mu - i g_X Q_X X_\mu,
\end{equation}
where $g_X$ and $Q_X$ is the gauge coupling constant and gauge charge of $U(1)_X$, respectively.
The scalar potential for the SM Higgs $H$ and $\varphi$ is given by
\begin{align}
V =  -\mu_H^2 H^\dagger H - \mu^2_\varphi \varphi^\ast \varphi + \frac{\lambda_H}{2} (H^\dagger H)^2 + \frac{\lambda_\varphi}{2} (\varphi^\ast \varphi)^2 + \lambda_{H \varphi} (H^\dagger H)(\varphi^\ast \varphi).
\label{eq:scalar-potential}
\end{align}
In the following discussion, we assume that $\mu_H^2$ and $\mu_\varphi^2$ are positive. 
The Lagrangians for the fermion and scalar DM candidate are given by 
\begin{subequations}
\begin{align}
\mathcal{L}^\chi_{DM} &=  \bar{\chi} ( i \Slash{D} -M_\chi) \chi + \left(y_L \overline{\chi_L^c} \chi_L \varphi + y_R \overline{\chi_R^c} \chi_R \varphi + h.c.\right), \\
\mathcal{L}^S_{DM} &= (D^\mu S)^\ast (D_\mu S) - M^2_S S^\ast S - \mu (\varphi S^\ast S^\ast + c.c.) \nonumber \\
& \quad - \lambda_S (S^\ast S)^2 - \lambda_{HS} (S^\ast S)(H^\dagger H) 
 - \lambda_{\varphi S} (\varphi^* \varphi)(S^\ast S),
\end{align}
\end{subequations}
where the superscript $c$ denotes charge conjugation of field, and the subscript $L$ and $R$ are left 
and right-handed chirality.

\subsection{Scalar Boson} \label{subsec:scalar-sector}
After $H$ and $\varphi$ develop a vacuum expectation value (VEV), $v/\sqrt{2}$ and $v_\varphi/\sqrt{2}$, 
respectively, the $U(1)_X$ and 
electroweak symmetries are spontaneously broken. Then, two physical CP-even scalar bosons remain 
in the spectrum as a mixture of the real parts of $H$ and $\varphi$.

Denoting the real parts as $\tilde{h}$ and $\tilde{\phi}$, the CP-even scalar bosons in mass eigenstate, 
$h$ and $\phi$, are expressed as
\begin{align}
\begin{pmatrix}
h \\
\phi
\end{pmatrix}
&=
U 
\begin{pmatrix}
\tilde{h} \\
\tilde{\phi}
\end{pmatrix}, 
\label{eq:scalar-eigenstates} 
\end{align}
where the diagonalization matrix $U$ and mixing angle $\alpha$ are defined by
\begin{subequations}
\begin{align}
U &=
\begin{pmatrix}
\cos\alpha & -\sin\alpha \\
\sin\alpha & \cos\alpha
\end{pmatrix}, 
\label{eq:diag-mat} \\
\tan 2\alpha &= \frac{2 \lambda_{H \varphi} v v_\varphi}{\lambda_H v^2 - \lambda_\varphi v_\varphi^2}. \label{eq:scalar-mixing}
\end{align}
\end{subequations} 
The masses of $h$ and $\phi$ are given by 
\begin{subequations}
\begin{align}
m_h^2 &= \lambda_H v^2 c_\alpha^2 + \lambda_\varphi v_\varphi^2 s_\alpha^2 
	+ 2 \lambda_{H\varphi} v v_\varphi s_\alpha c_\alpha, \\
m_\phi^2 &= \lambda_\varphi v_\varphi^2 s_\alpha^2 + \lambda_H v^2 s_\alpha^2 
	- 2 \lambda_{H\varphi} v v_\varphi s_\alpha c_\alpha.
\end{align}
\label{eq:scalar-masses}
\end{subequations}
Note that $h$ becomes the SM Higgs boson in the limit of $\alpha \to 0$.
The scalar boson $\phi$ can interact with the SM fermions and weak gauge bosons through the mixing. The interaction Lagrangian is 
given by
\begin{align}
\label{eq:int-f-phi}
\mathcal{L}^{\mathrm{SM}}_{\phi\mathrm{-int}} &= \sum_{f} \frac{m_f}{v} \sin \alpha \phi \bar f f
+ \frac{2 m_W^2}{v} \sin \alpha \phi W^+_\mu W^{-\mu}+ \frac{m_Z^2}{v} \sin \alpha \phi Z_\mu Z^\mu ,
\end{align}
where $f$ runs over the SM fermions. Assuming $\epsilon \ll 1$, the neutral weak gauge boson $Z$ and 
its mass $m_Z$ are approximated by that of the SM while the charged one $W^\pm$ 
is exactly the same as that of the SM. The interactions of $\phi$ with the dark matter are given in Sec. \ref{subsec:DM-sector}.

\subsection{Dark Photon} \label{subsec:gauge-sector}
The gauge bosons in our model acquire masses after the spontaneous symmetry breakings.
The electrically neutral components of the gauge bosons mix each other through off-diagonal masses and 
the kinetic mixing while the charged ones remain the same as those of the SM. 
Assuming $\epsilon \ll 1$, new gauge field $X$ is approximately identified as mass eigenstate and we denote it as dark photon $A'$ hereafter.  
The mass of $A'$ is expressed as
\begin{align}
m_{A'} = g_X v_\varphi.
\end{align}
The gauge interaction of dark photon with the SM particles and $\phi$ is given by 
\begin{align}
\label{eq:int-gauge}
\mathcal{L}^{\mathrm{SM}}_{A'\mathrm{-int}} &=
 e \epsilon \cos \theta_W J_{\rm EM}^\mu A'_\mu  + g_X m_{A'} \cos \alpha \phi A'_\mu A'^\mu,
\end{align}
where $\theta_W$ is the Weinberg angle, and $e$ and $J_{\rm EM}^\mu$ are the elementary charge and electromagnetic currents of the SM.
The gauge interactions of the dark photon to the DM are given in next subsection.

\subsection{Dark Matter} \label{subsec:DM-sector}
In this subsection, we discuss mass eigenstates and interactions of our fermion and scalar DM candidates 
after the spontaneous symmetry breakings. \\
{\bf (I) Fermion DM} \\ 
The mass terms of $\chi$ are given by %
\begin{equation}
\mathcal{L}_{M_\chi} = M_\chi (\bar \chi_L \chi_R + \bar \chi_R \chi_L) + \left( \frac{y_L v_\varphi}{\sqrt{2}} \overline{\chi_L^c} \chi_L + \frac{y_R v_\varphi}{\sqrt{2}} \overline{\chi_R^c} \chi_R + h.c. \right).
\end{equation}
Thus we can rewrite these terms in the basis of $(\chi_L, \chi^c_R)^T$ such that 
\begin{equation}
\mathcal{L}_{M_\chi} = \frac{1}{2}
( \overline{ \chi^c_L}~ \overline{ \chi_R} )
 \begin{pmatrix} m_L & M_\chi \\ M_\chi & m_R \end{pmatrix} 
\begin{pmatrix} \chi_L \\ \chi_R^c \end{pmatrix}+ h.c.\, ,
\end{equation}
where $m_{L(R)} \equiv \sqrt{2} y_{L(R)} v_\varphi$. Due to the diagonal elements of the mass matrix, 
$\chi$ is decomposed 
into two mass eigenstates. 
The mass eigenvalues are obtained by diagonalizing the matrix such that
\begin{equation}
m_{\chi_1,\chi_2} = \frac{m_L + m_R}{2} \pm \frac12 \sqrt{(m_L - m_R)^2 + 4 M^2},
\end{equation}
where we chose $m_{\chi_1} < m_{\chi_2}$ as convention.
Mass eigenstates $\chi_1$ and $\chi_2$ are also given by
\begin{subequations}
\begin{align}
& \begin{pmatrix} \chi_1 \\ \chi_2 \end{pmatrix} = 
\begin{pmatrix} \cos \theta_\chi & - \sin \theta_\chi \\ \sin \theta_\chi & \cos \theta_\chi \end{pmatrix}
\begin{pmatrix} \chi_L \\ \chi_R^c \end{pmatrix}, \\
& \tan 2 \theta_\chi = \frac{2 M_\chi}{m_L - m_R}.
\end{align}
\end{subequations}
The gauge interactions among the mass eigenstates and $A'$ can be written by
\begin{align}
\mathcal{L}^{\chi}_{A'\mathrm{-int}}
=  g_X A'_\mu \left[ \cos 2 \theta_\chi (\bar \chi_1 \gamma^\mu \chi_1 - \bar \chi_2 \gamma^\mu \chi_2) +  \sin 2 \theta_\chi  (\bar \chi_1 \gamma^\mu \chi_2 + \bar \chi_2 \gamma^\mu \chi_1) \right], 
\end{align}
and the interaction to $\phi$ and $h$ are 
\begin{align}
\mathcal{L}^{\chi}_{\phi\mathrm{-int}}
= & \frac{1}{\sqrt{2}} y_L (c_\alpha \phi - s_\alpha h)(c_\chi^2 \overline{\chi^c_1} \chi_1 +  c_\chi s_\chi (\overline{\chi^c_1} \chi_2 + \overline{ \chi_1} \chi^c_2) + s^2_\chi \overline{\chi^c_2} \chi_2) \nonumber \\
& + \frac{1}{\sqrt{2}} y_R (c_\alpha \phi - s_\alpha h)(s_\chi^2 \overline{ \chi^c_1} \chi_1 -  c_\chi s_\chi (\overline{\chi^c_1} \chi_2 + \overline{\chi_1} \chi^c_2) + c^2_\chi \overline{\chi^c_2} \chi_2) + h.c.,
\end{align}
where $s_\theta(c_\theta)$ stands for $\sin \theta_{\chi} (\cos \theta_{\chi})$.
\\

\noindent
{\bf (II) Scalar DM } \\ 
In the case of the scalar DM, the mass terms are given by
\begin{subequations}
\begin{align}
& \mathcal{L}_{M_S} = \frac12 \left( \tilde M_S^2 + \sqrt{2} \mu v_\varphi  \right) s^2 +\frac12 \left(\tilde M_S^2 - \sqrt{2} \mu v_\varphi  \right) a^2, \\
& \tilde M_S^2 \equiv M_S^2 + \frac{\lambda_{HS}}{2} v^2 + \frac{\lambda_{\varphi S}}{2} v_\varphi^2.
\end{align}
\end{subequations}
Thus the CP-even $s$ and CP-odd $a$ components have different masses given by
\begin{equation}
m^2_{s,a} = \tilde M^2_S \pm \sqrt{2} \mu v_\varphi
\end{equation}
where $+$ and $-$ in RHS stand for $s$ and $a$ respectively.
The interaction terms among $A'$ and $s,~a$ are written by
\begin{equation}
\mathcal{L}^{S}_{A'\mathrm{-int}} = \frac{1}{2} g_X A'_\mu (s \partial^\mu a - a \partial^\mu s  ) 
+ \frac{1}{8} g_X^2 A'_\mu A'^\mu (s^2 + a^2).
\end{equation}
In addition scalar interactions from the potential are given by 
\begin{equation}
\mathcal{L}^{S}_{\phi\mathrm{-int}} =  -\frac{\mu}{\sqrt2} \tilde \phi (s^2 - a^2) - \frac{\lambda_{HS}}{4} (\tilde h^2 + 2 v \tilde h)(s^2+a^2) - \frac{\lambda_{\varphi S}}{4} (\tilde \phi^2 + 2 v_\varphi \tilde \phi)(s^2 +a^2),
\end{equation}
where $\tilde \phi$ and $\tilde h$ are written by mass eigenstate by \eqref{eq:scalar-eigenstates}.

\section{Analysis} \label{sec:analysis}

In this section we discuss the production of the excited states through the decay of the scalar boson 
$\phi$ at the LHC. 
In our notation, $\chi_2(a)$ and $\chi_1(s)$ is the excited state and DM for the fermion(scalar) DM case.
We then estimate the number of events for the decay of the excited state at the FASER experiment.

\subsection{Decay widths}

Here we discuss decay processes of our new particles; $\{\phi, \chi_2 \}$ for fermion DM case and $\{ \phi, a \}$ for scalar DM case.
We adopt mass relation $m_{\chi_2(a)} - m_{\chi_1(s)} < m_{A'}$ so that $\chi_2(a)$ dominantly decay into $\chi_2[a] \to A'^{*}(\to \bar f f) \chi_1[s]$ where $f$ is the SM fermions.
In addition we require $m_\phi > 2 m_{\chi_2(a)}$ to produce $\chi_2(a)$ pair through the scalar 
boson decay at the LHC.

\noindent
{\bf (I) Fermion DM} \\ 
The decay widths for $\phi \to \chi_i \chi_j$ are obtained such that
\begin{subequations}
\begin{align}
& \Gamma_{\phi \to \chi_1 \chi_1} = \frac{c_\alpha^2}{8 \pi} (y_L c_\chi^2 + y_R s_\chi^2)^2 m_\phi \left( 1 - \frac{4 m_{\chi_1}^2}{m_\phi^2} \right)^{\frac32}, \\
& \Gamma_{\phi \to \chi_2 \chi_2} = \frac{c_\alpha^2}{8 \pi} (y_L s_\chi^2 + y_R c_\chi^2)^2 m_\phi \left( 1 - \frac{4 m_{\chi_2}^2}{m_\phi^2} \right)^{\frac32}, \\
& \Gamma_{\phi \to \chi_1 \chi_2} = \frac{c_\alpha^2 s_\chi^2 c_\chi^2}{4 \pi} (y_L - y_R )^2 m_\phi \lambda \left(1, \frac{ m_{\chi_1}^2}{m_\phi^2}, \frac{ m_{\chi_2}^2}{m_\phi^2} \right)
 \left( 1 - \frac{(m_{\chi_1}+m_{\chi_2})^2}{m_\phi^2} \right), 
\end{align}
\end{subequations}
where 
\begin{align}
    \lambda(x,y,z) = \sqrt{(x-y-z)^2 - 4yz}. \label{eq:kallen-f}
\end{align}
Note that $\phi$ can decay into the SM particles via Higgs mixing. For simplicity, we assume $s_\alpha \ll 1 $ 
so that these decays are  highly suppressed and ignore these decay modes in our analysis.

We also consider $A'$ decay processes. 
It can decay into the dark fermions $\chi_{1,2} \chi_{1,2}$ and/or $\chi_1 \chi_2$ when these modes are 
kinematically allowed.
The decay widths of these modes are estimated as 
\begin{subequations}
\begin{align}
\Gamma_{A' \to \chi_{1(2)} \chi_{1(2)}} &= \frac{g_X^2 c_{2\chi}^2}{48 \pi} m_{A'} \left( 1 - \frac{4 m_{\chi_{1(2)}}^2}{m_{A'}^2} \right)^{\frac32}, \\
\Gamma_{A' \to \chi_1 \chi_2} &= \frac{g_X^2 s_{2\chi}^2}{24 \pi} m_{A'} \lambda_{A'} \nonumber \\ 
&\times \left[1 - \frac{m_{\chi_1}^2}{m_{A'}^2} - \frac{m_{\chi_2}^2}{m_{A'}^2} - 6 \frac{m_{\chi_1}^2m_{\chi_2}^2}{m_{A'}^4} +
\sqrt{ \left(\lambda_{A'}^2 +\frac{4m_{\chi_2}^2}{m_{A'}^2} \right) \left(\lambda_{A'}^2 + \frac{4m_{\chi_1}^2}{m_{A'}^2} \right)} \right],
\end{align}
\end{subequations}
where $s_{2\chi}(c_ {2 \chi}) \equiv \sin 2 \theta_\chi (\cos 2 \theta_\chi)$ and $\lambda_{A'} = \lambda(1,m^2_{\chi_1}/m_{A'}^2,m^2_{\chi_2}/m_{A'}^2)$. 
Our $A'$ also decays into the SM fermions through the kinetic mixing. 
The partial width of $A' \to f \bar f$ decay is given by
\begin{subequations}
\begin{align}
\Gamma_{A' \to f \bar f} & = \frac{\epsilon^2 e^2}{12 \pi} m_{A'} \left( 1 - \frac{4m_f^2}{m_{A'}^2} \right)^{\frac12} \left( 1 + \frac{2 m_f^2}{m_{A'}^2} \right), \\
\Gamma_{A' \to {\rm hadrons}} & = \Gamma_{A' \to \mu^+ \mu^-} R(s = m^2_{A'}),
\end{align}
\end{subequations}
where $R(s) \equiv \sigma(e^+ e^- \to {\rm hadrons})/\sigma(e^+ e^- \to \mu \bar \mu)$ with $s$ being the center of mass energy.

\begin{figure}[t!]\begin{center}
\includegraphics[width=80mm]{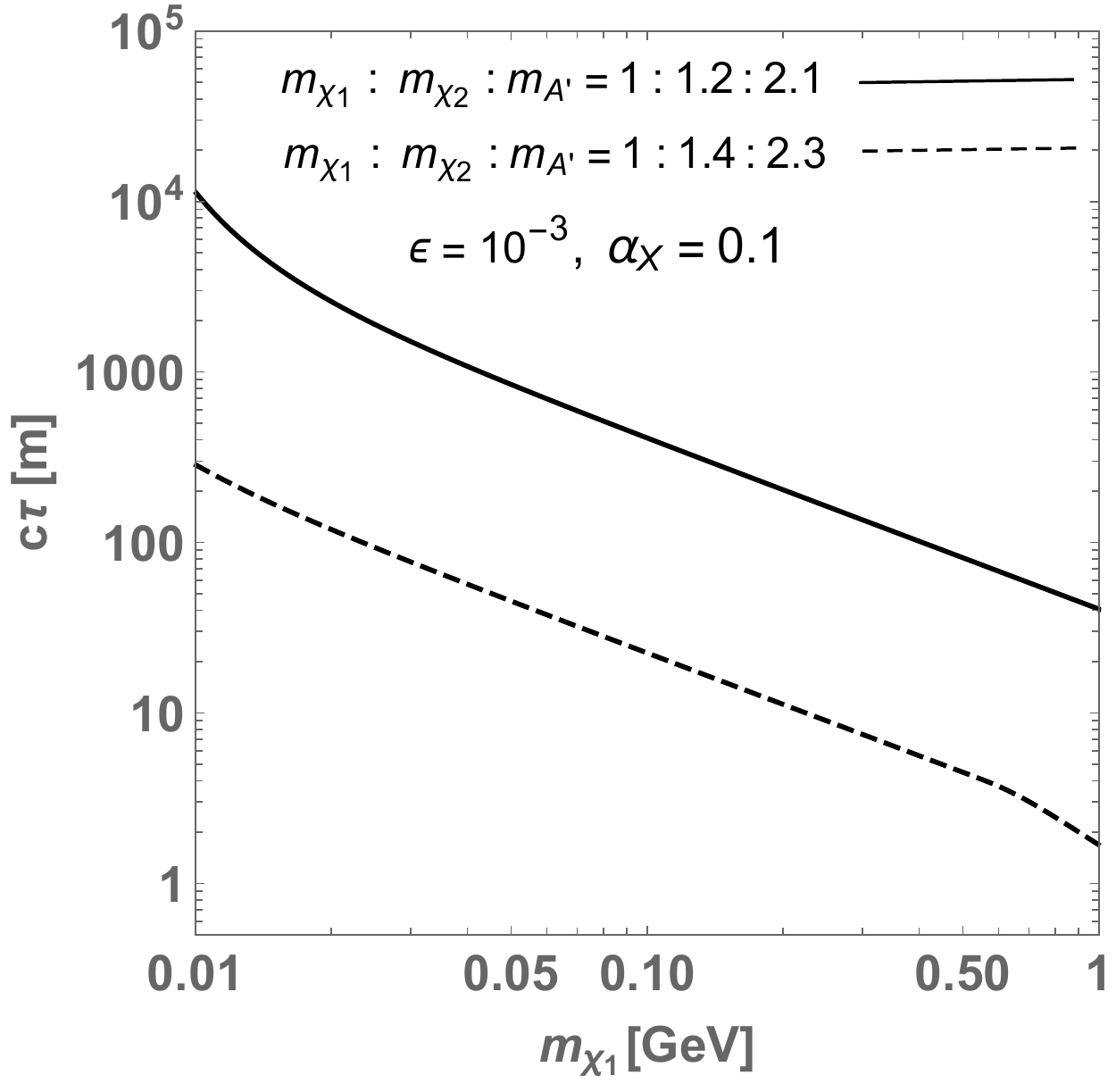}
\includegraphics[width=80mm]{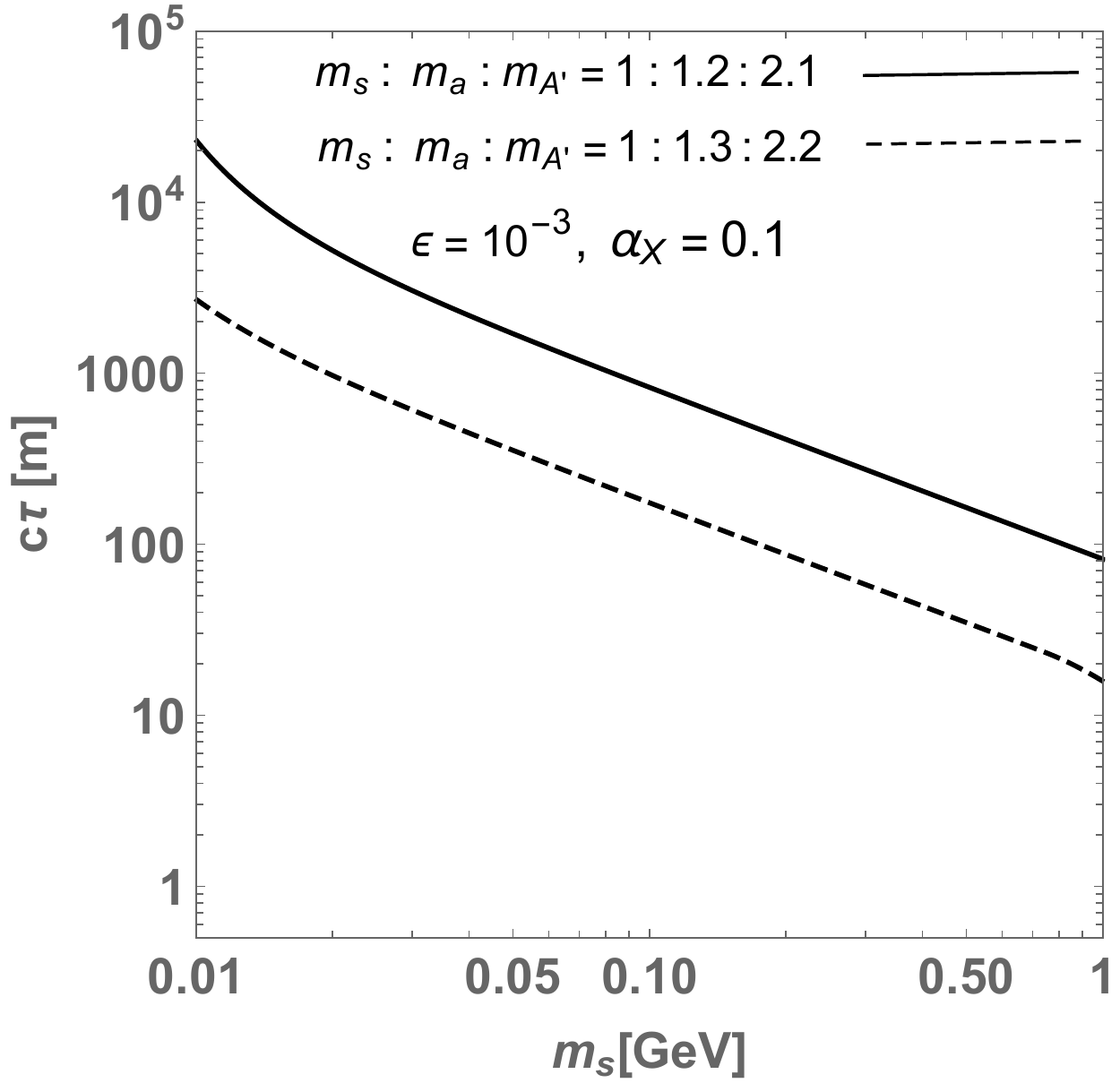} 
\caption{Total decay length of $\chi_2$ and $a$ as functions of $m_{\chi_1}$ and $m_s$ given in the left and right plots.}   
\label{fig:width1}
\end{center}
\end{figure}

The decay width of $\chi_2$ is given by 
\begin{subequations}
\begin{align}
& \Gamma_{\chi_2} =  \frac{g_X^2 s_{2\chi}^2 e^2 \epsilon^2 c_W^2}{256 \pi^3 m^3_{\chi_2}} \int_{s_2^-}^{s_2^+} d s_2 \int_{s_1^-}^{s_1^+} d s_1
\frac{1}{(m^2_{\chi_1} + m^2_{\chi_2} + 2 m^2_e - s_1 - s_2 - m^2_{A'})^2 + m^2_{A'} \Gamma^2_{A'} } \nonumber \\
& \qquad \qquad \qquad \times \bigl[ - s_1^2 - s_2^2 + (s_1+s_2)(m^2_{\chi_1} + m^2_{\chi_2} +4 m^2_e ) - 2 m^2_{\chi_1} m^2_{\chi_2} -  2 m^2_{\chi_1} m^2_e \nonumber \\
& \qquad  \qquad \qquad  \qquad - 2 m^2_{\chi_2} m^2_e- 6 m^4_e - 2 m_{\chi_1} m_{\chi_2} (m^2_{\chi_1} + m^2_{\chi_2} + 4 m_e^2 - s_1 - s_2) \bigr] \nonumber \\
& \qquad \qquad \qquad \times \frac{1}{BR(A' \to e^+ e^-)_{m_{A'^*} = m_{ee}}}, \\
& s_1^\pm =   m^2_{\chi_1} + m^2_e \nonumber \\
& \qquad + \frac{1}{2 s_2} [ (m^2_{\chi_2} - m_e^2 - s_2)(m^2_{\chi_1} - m_e^2 + s_2) \pm \lambda(s_2,m^2_{\chi_2},m^2_e) \lambda(s_2,m^2_{\chi_1},m^2_e)] \label{eq:s1} \\ 
& s_2^- = (m_{\chi_1}+ m_e)^2, \quad s_2^+ = (m_{\chi_2} - m_e)^2, \label{eq:s2}
\end{align} 
\end{subequations}
where $s_{2\chi} \equiv \sin 2 \theta_\chi$ and $BR(A' \to e^+ e^-)_{m_{A'^*} = m_{ee}}$ is the branching ratio of a $A' \to e^+ e^-$ mode with $m_{ee}$ being invariant mass of electron-positron pair; $m_{ee}^2 = m_{\chi_1}^2+m^2_{\chi_2}+2m^2_e - s_1 -s_2$.
The left plot of Fig.~\ref{fig:width1} shows the total decay length of $\chi_2$ as a function of $m_{\chi_1}$ where we fix mass ratio as $m_{\chi_1}: m_{\chi_2}: m_{A'} = 1:1.2:2.1(1:1.4:2.3)$ and 
other parameters as $\epsilon = 10^{-3}$ and $\alpha_X \equiv g_X^2/4\pi = 0.1$.
\\

\noindent
{\bf (II) Scalar DM} \\ 
The decay widths for $\phi \to aa(ss)$ are obtained such that
\begin{subequations}
\begin{align}
& \Gamma_{\phi \to ss} = \frac{c_\alpha^2}{32 \pi} C_{\phi ss}^2 m_\phi \left( 1 - \frac{4 m_{s}^2}{m_\phi^2} \right)^{\frac12}, \\
& \Gamma_{\phi \to aa} = \frac{c_\alpha^2}{32 \pi} C_{\phi aa}^2 m_\phi \left( 1 - \frac{4 m_{a}^2}{m_\phi^2} \right)^{\frac12}, 
\end{align}
\end{subequations}
where the coefficient $C_{\phi ss(aa)}$ is given by
\begin{subequations}
\begin{align}
C_{\phi ss} = -\frac{1}{\sqrt2} \mu c_\alpha - \frac12 \lambda_{HS} v s_\alpha - \frac12 \lambda_{\varphi S} v_\varphi c_\alpha, \\
C_{\phi aa} = \frac{1}{\sqrt2} \mu c_\alpha - \frac12 \lambda_{HS} v s_\alpha - \frac12 \lambda_{\varphi S} v_\varphi c_\alpha.
\end{align}
\end{subequations}
As mentioned above, we ignore decay of $\phi$ into the SM particles. 

In the scalar DM case $A'$ cannot decay into $ss$ or $aa$ mode while $A' \to sa$ mode is possible if kinematically allowed.
The decay width of the mode is given by
\begin{equation}
\Gamma_{A' \to s a} = \frac{g_X^2}{24 \pi} m_{A'} \left[ \lambda \left(1, \frac{m_s^2}{m_{A'}^2}, \frac{m_a^2}{m_{A'}^2} \right) \right]^3.
\end{equation}
In our analysis, we choose $m_a + m_s > m_{A'}$ and $A'$ only decays into SM fermions through the kinetic mixing.

As in the fermion DM case, the decay width for $a$ is given by
\begin{align}
& \Gamma_{a} =  \frac{g_X^2 e^2 \epsilon^2 c_W^2}{256 \pi^2 m^2_{\chi_2}} \int_{s_2^-}^{s_2^+} d s_2 \int_{s_1^-}^{s_1^+} d s_1
\frac{1}{(m^2_{\chi_1} + m^2_{\chi_2} + 2 m^2_e - s_1 - s_2 - m^2_{A'})^2 + m^2_{A'} \Gamma^2_{A'} } \nonumber \\
& \qquad \qquad \qquad \times \bigl[ (s_1 - m_e^2)(s_2 - m_e^2) - m_s^2 m_a^2 \bigr] \nonumber \\
& \qquad \qquad \qquad \times \frac{1}{BR(A' \to e^+ e^-)_{m_{A'^*} = m_{ee}}}, 
\end{align}
where $s_1^\pm$, $s_2^\pm$ and $m_{ee}$ are obtained by substituting $m_{\chi_1}$ and $m_{\chi_2}$ into $m_s$ and $m_a$ in \eqref{eq:s1} and \eqref{eq:s2}.
The right plot of Fig.~\ref{fig:width1} shows the total decay length of $a$ as a function of $m_{s}$ where we fix mass ratio as $m_{s}: m_{a}: m_{A'} = 1:1.2:2.1(1:1.3:2.3)$ and other parameters the same as fermion DM case.

\subsection{DM annihilation and relic density}

In this subsection we discuss annihilation processes of our DM candidates and its relic density where DM mass scale is $m_{DM} \lesssim \mathcal{O}(1)$ GeV.
Particles in dark sector can interact with the SM ones via scalar and $A'$ exchange through Higgs mass mixing or gauge kinetic mixing effects.
In our scenario, we chose the parameters so that $A'$ exchanging processes are dominant. \\

\noindent
{\bf (I) Fermion DM} \\
In this case we have following annihilation processes via $A'$ exchange 
\begin{equation}
\chi_1 \chi_1 \to A' \to \bar f f, \quad \chi_1 \chi_2 \to A' \to \bar f f,
\end{equation}
where the second process is coannihilation process.
In our analysis below, we chose mixing angle for $\chi_{1,2}$ by $\theta_\chi = \pi/4 - 0.05$ that is close to maximal mixing  $\sin \theta_\chi = \cos \theta_\chi = 1/\sqrt{2}$ corresponding to $y_L \simeq y_R$.
We also chose $m_{\chi_1} \simeq m_{\chi_2}$ and $m_{\chi_1} + m_{\chi_2} \simeq m_{A'}$ so that the (co)annihilation cross section is enhanced.
In this choice both annihilation and coannihilation processes contribute to Boltzmann equation in estimating relic density of fermionic DM.

\noindent
{\bf (II) Scalar DM} \\
In this case we have following annihilation process via $A'$ exchange 
\begin{equation}
s a \to A' \to \bar f f, 
\end{equation}
where we have only coannihilation process.
We also chose $m_{s} + m_{a} \simeq m_{A'}$ so that the coannihilation cross section is enhanced. Note that we do not have annihilation process $ss(aa) \to A' \to \bar f f$ in contrast to fermionic DM case.
As we will see, the correct relic abundance requires relatively large kinetic mixing mainly due 
to the absence of the annihilation process.

We implement relevant interactions in {\it micrOMEGAs 5}~\cite{Belanger:2014vza} to estimate relic density for both fermion and scalar DM cases.

\subsection{Scalar boson production from meson decay}
The scalar boson, $\phi$, is mainly produced in the decays of mesons through the mixing with 
the SM Higgs, \eqref{eq:scalar-mixing}.
For the small scalar mixing angle, the decay branching ratios of the meson into $\phi$ and a lighter meson 
are given by ~\cite{Feng:2017vli}
\begin{subequations}
\begin{align}
\label{eq:br-sigma-prod-B}
   {\rm Br}(B \to X_s \phi) &\simeq 5.7 \left( 1 - \frac{m_\phi^2}{m_b^2} \right)^2 \alpha^2~, \\
   {\rm Br}(K^\pm \to \pi^\pm \phi) &= 2.0 \times 10^{-3} \frac{2 p_\phi^0}{m_{K^\pm}} \alpha^2~, \\
   {\rm Br}(K_L \to \pi^0 \phi) &= 7.0 \times 10^{-3} \frac{2 p_\phi^0}{m_{K^0_L}} \alpha^2~, \\
   {\rm Br}(K_S \to \pi^0 \phi) &= 2.2 \times 10^{-6} \frac{2 p_\phi^0}{m_{K^0_S}} \alpha^2~,  
\end{align}
\end{subequations}
where $p_\phi^0 = \lambda^{1/2}(m_{K}^2, m^2_{\pi}, m_\phi^2) /(2 m_{K})$ is the three-momentum of the scalar boson 
in the parent meson's rest frame.
%
\begin{figure}[t]
\centering
\includegraphics[width=0.9\textwidth]{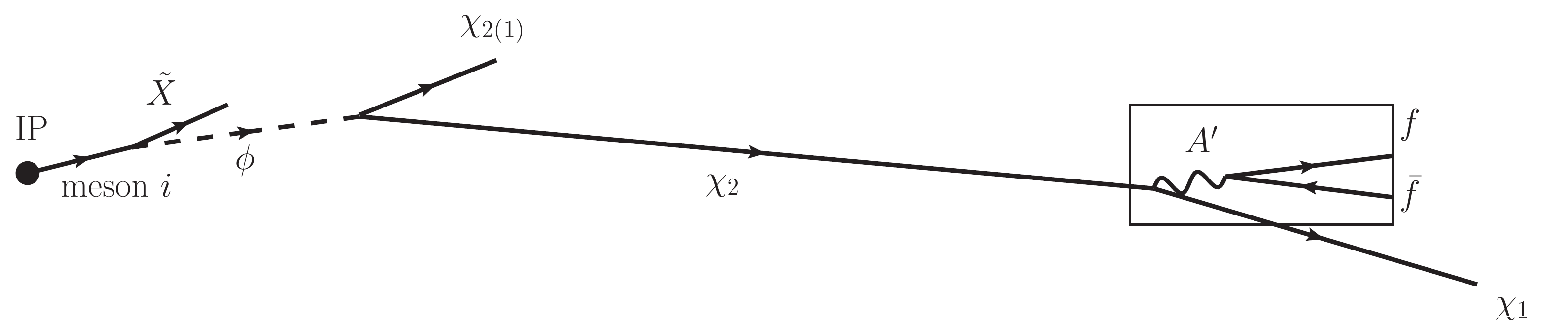}
\caption{
Sketch of the production of $\chi_2$ from the decay of the scalar boson, and the subsequent decay into $\chi_1$ and $f$-$\bar{f}$ pair. IP denotes the Interaction Point at the ATLAS detector. 
}
\label{fig:process}
\end{figure}
In our setup, the produced $\phi$ boson dominantly decays into the dark particles, $\chi_1,\chi_2$ or 
$s,a$. 
Then, the excited dark particles $\chi_2$ and $a$ decay into the dark matter $\chi_1$ and $s$, 
respectively, in the detector. We depict this processes in Figure \ref{fig:process} for the fermion inelastic dark matter 
case.
In this case, the expected number of the decays of $\chi_2$ is estimated by
\begin{align}
    \label{eq:num-of-event-f}
N_{\chi_2} &= \mathcal{L} \sum_{i:{\rm meson}} \int dp_i d\theta_i \int d\bm{p}_{\chi_2} \int d\bm{p}_\phi
   \frac{d\sigma_{pp \to i X}}{dp_i d\theta_i } {\rm Br}(i \to \tilde{X} \phi) \nonumber \\   
   &\qquad \times \bigg[ \sum_{j=1,2} {\rm Br}(\phi \to (\chi_{2})_1 (\chi_{2})_2 ) \mathcal{P}_{(\chi_{2})_j}^{\rm det}(\bm{p}_{(\chi_2)_j}, \bm{p}_\phi)
   + {\rm Br}(\phi \to \chi_{1} \chi_{2}) \mathcal{P}_{\chi_{2}}^{\rm det}(\bm{p}_{\chi_2}, \bm{p}_\phi) \bigg], 
\end{align}
where $\frac{d\sigma_{pp \to i X}}{dp_i d\theta_i }$ is the differential cross section of $\phi$ production 
at the ATLAS, and $\mathcal{P}_{\chi_{2}}^{\rm det}(\bm{p}_{\chi_2},\bm{p}_\phi)$ is the decay probability of 
$\chi_2$ taking the decay position of $\phi$ into account. 
Label $j$ distinguishes two $\chi_2$ produced from the $\phi$ decay.  
We consider the contributions from the decays of charged and neutral long/short-lived kaons, as well as $b$-quarks. 
The light meson spectra are simulated by \texttt{EPOS-LHC}~\cite{Pierog:2013ria} as implemented in the package \texttt{CRMC}~\cite{crmc:html}, while the $b$-quark spectrum is generated by \texttt{Pythia8}~\cite{Sjostrand:2007gs} with A3 tune~\cite{Skands:2014pea,ATLAS:2016puo}.
Our result is in good agreement with \texttt{FORESEE}~\cite{Kling:2021fwx}. 
In the scalar boson case, the expected number of the decays of $a$ is given by
\begin{align}
    \label{eq:num-of-event-s}
N_s &= \mathcal{L} \sum_{i:{\rm meson}} \int dp_i d\theta_i \int d\bm{p}_{a} \int d\bm{p}_\phi
   \frac{d\sigma_{pp \to i X}}{dp_i d\theta_i } {\rm Br}(i \to \tilde{X} \phi) \nonumber \\   
   &\qquad \times \sum_{j=1,2} {\rm Br}(\phi \to (a)_1 (a)_2 ) \mathcal{P}_{a_j}^{\rm det}(\bm{p}_{a_j},\bm{p}_\phi),
\end{align}
where $\mathcal{P}_{a}^{\rm det}(\bm{p}_{a},\bm{p}_\phi)$ is the decay probability of $a$.
Following \cite{Araki:2020wkq}, the decay probability of the particle $i$ $(=\chi_2$ or $a)$ at the FASER detector is given by
\begin{align}
\label{eq:prob}
\mathcal{P}_{i}^{\rm det}(\bm{p}_{i}, \bm{p}_\phi) &= 
\frac{1}{\bar{d}_\phi \cos\theta_\phi} \int_{z_{\phi,\mathrm{min}}}^{z_{\phi,\mathrm{max}}} dz_\phi e^{-\frac{z_\phi}{\bar{d}_\phi \cos\theta_\phi}}
\frac{1}{\bar{d}_{i} \cos\theta_{i}} \int_{z_{i,\mathrm{min}}}^{L_{\mathrm{max}}} dz_{i} e^{-\frac{z_{i}-z_\phi}{\bar{d}_{i} \cos\theta_{i}}} 
\nonumber \\
&\quad \times \Theta(R - r_{i,R}) \Theta(R - r_{i,F} ),
\end{align}
where $\bm{p}_{i(\phi)}$ and $\theta_{i(\phi)}$ denote the momentum and angle with respect to 
the beam axis ($z$-axis), respectively, and $\bar{d}_{i(\phi)}$ denotes the decay length of 
the excited state (the scalar boson) in the laboratory frame. 
\begin{table}[t]
\begin{tabular}{|c|c|c|c|c|} \hline
 \hspace{2cm} & ~~~$L_{\rm min}$~(m)~~~ & ~~~$L_{\rm max}$~(m)~~~ & ~~~$R$~(m)~~~ & ~~~$\mathcal{L}$~(ab$^{-1}$)~~~ \\ \hline \hline 
 FASER  & 478.5 & 480 & 0.1 & 0.15 \\ \hline
 FASER~2 & 475 & 480 & 1.0 & 3.0 \\ \hline
\end{tabular}
\caption{
Dimension of the FASER detector and integrated luminosity used in this study. 
$L_{\rm min}$ and $L_{\rm max}$ are the distance to the front and rear end of the FASER detector from the IP, respectively. 
$R$ is detector radius, respectively. $\mathcal{L}$ is the integrated luminosity.
}
\label{tab:faser-dimension}
\end{table}
The detector radius and the distance from the IP to the rear (front) end of the FASER detector are denoted as 
$R$ and $L_{\rm max (min)}$, respectively, which are given in Table \ref{tab:faser-dimension}.
In the step functions, $r_{i,R}$ and $r_{i,F}$ are the distance of the particle $i$ from the beam axis at $z_{i} = L_{\mathrm{max}}$ and $L_{\mathrm{min}}$, respectively. 
The integral regions can be determined so that the step functions are satisfied 
for given $\bm{p}_{i}$ and $\bm{p}_\phi$. See \cite{Araki:2020wkq} for the details of the calculation of the decay probability.

\subsection{Signal event at FASER}

In the end of this section, we show our results of the expected number of signal events 
at the FASER experiment. 
For the fermion(scalar) DM case, we scan the DM mass and kinetic mixing parameter as follows;
\begin{equation}
m_{\chi_1 (s)} \in [0.001, 1] \ {\rm GeV}, \quad \epsilon \in [10^{-7}, 10^{-1}],
\end{equation}
where the gauge coupling are fixed to be $g_X^2/(4\pi) = 0.1$ so that small kinetic mixing is allowed 
consistent with the DM relic abundance. 
We adapt the following relations for $m_{\chi_2(a)}$ and $A'$ mass 
\begin{subequations}
\begin{align}
1.&~~m_{\chi_1(s)} : m_{\chi_2(a)} : m_{A'} = 1 : 1.2 : 2.1, \label{eq:spectrum-f}\\
2.&~~m_{\chi_1(s)} : m_{\chi_2(a)} : m_{A'} = 1 : 1.4(1.3) : 2.3(2.2) \label{eq:spectrum-s},
\end{align}
\label{eq:spectrum}
\end{subequations}
where we chose $m_{\chi_1(s)} + m_{\chi_2(a)} \sim m_{A'}$ to enhance the (co)annihilation cross 
section. In these spectra, $A'$ dominantly decays into the dark matter $\chi_1$ and $s$, and hence is invisible.
It should be noticed that the excited states cannot be produced from the on-shell $A'$ decay in the above two spectra. These particles can be produced directly through an off-shell $A'$ from $pp$ 
collision. The cross section of such processes scales as the inverse of the center of mass energy squared, and hence will be suppressed in the LHC experiment.\footnote{These will be studied in our 
future work.}
Therefore the scalar boson decay is the main source of $\chi_2$ and $a$ in the above spectra.
To examine the sensitivity from the scalar boson decay, we adopt the mass relation for $m_{\chi_1(s)}$ and $\phi$, and the scalar mixing angle 
\begin{align}
    m_{\chi_1(s)} : m_\phi = 1: 4, \\
    \alpha = 10^{-4}.
\end{align}
With these parameters, the expected number of the signal event $\chi_2 \to \chi_1 f \bar{f}$ or 
$a \to s f \bar{f}$ is calculated by \eqref{eq:num-of-event-f} or (\ref{eq:num-of-event-s}), respectively. For the FASER setup given in Table \ref{tab:faser-dimension}, we do not find 
viable sensitivity region and therefore only show the results for the FASER 2 setup.

\begin{figure}[t]
\centering
\includegraphics[width=0.48\textwidth]{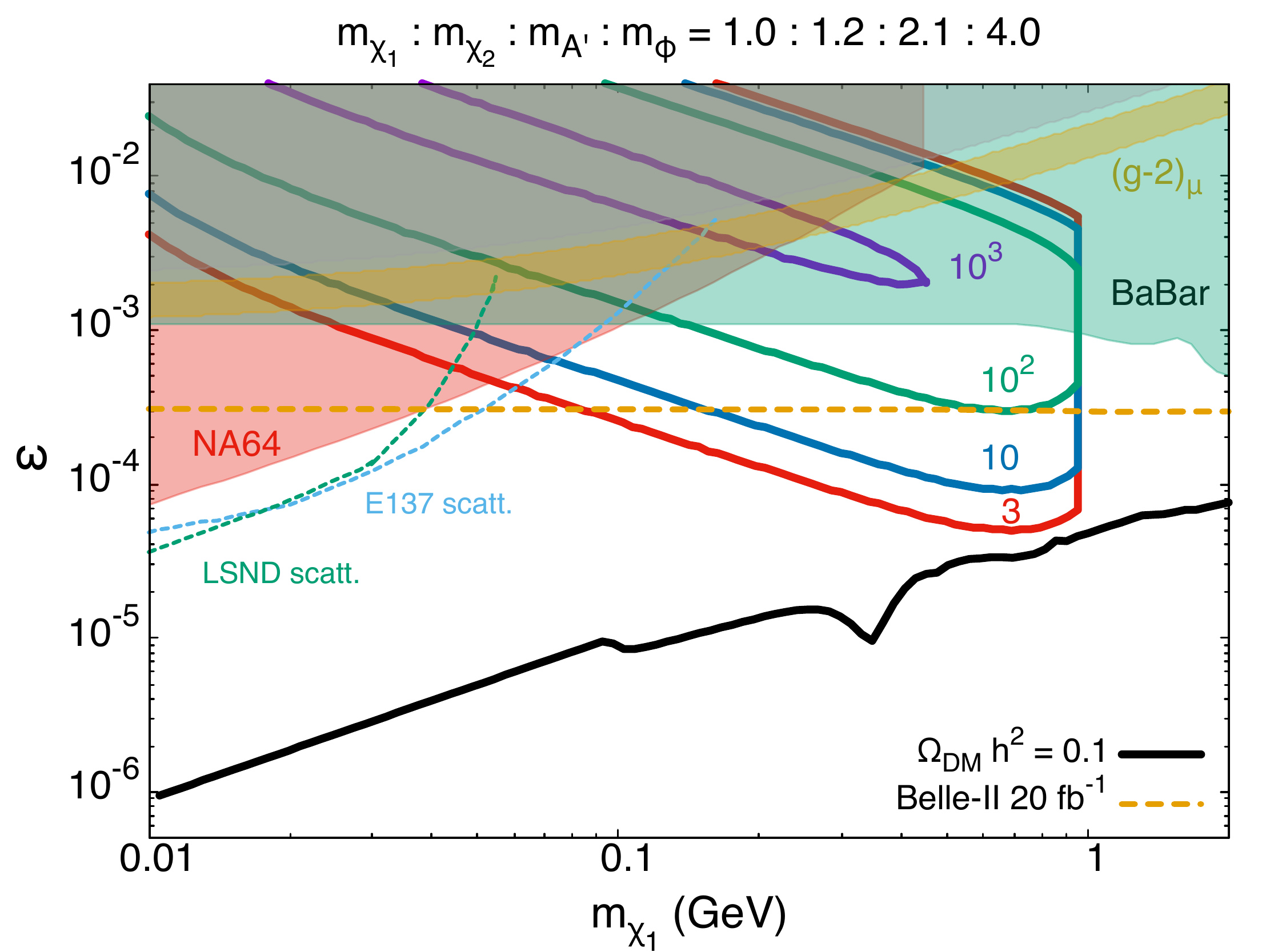} \quad
\includegraphics[width=0.48\textwidth]{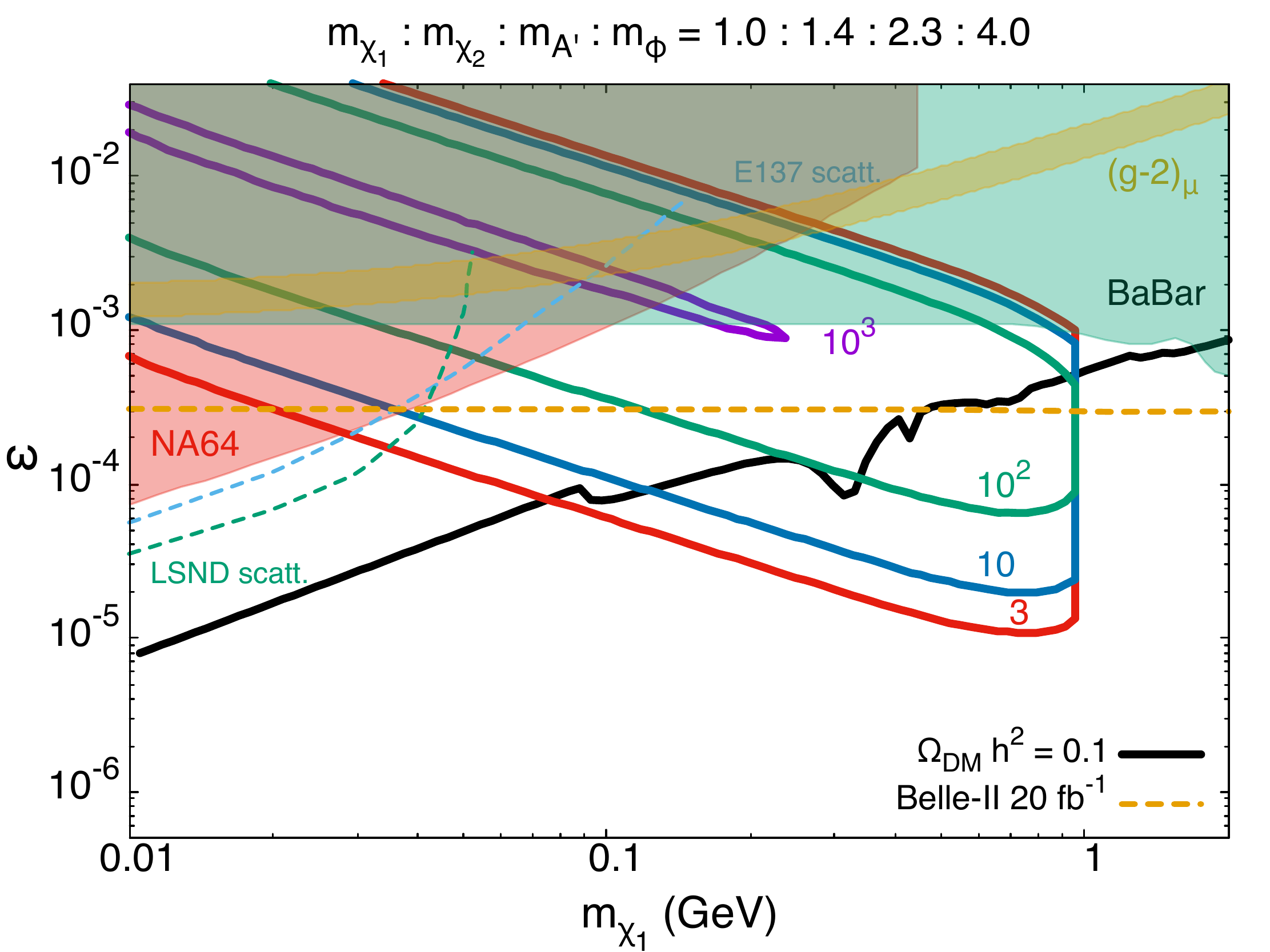}
\caption{
Sensitivity region of the fermion inelastic dark matter decay. Red, blue, green and purple curves represent contours of the expected number of the signal events $3,~10,~10^2$ and $10^3$, respectively.
Black curve represents the relic abundance of the dark matter. Yellow dashed line is the projection 
of Belle-II sensitivity \cite{Belle-II:2018jsg}. 
The filled color regions are excluded by NA64 (red) \cite{Banerjee:2019pds} and BaBar (green) \cite{BaBar:2017tiz}, respectively.
}
\label{fig:f-idm}
\end{figure}
Figure \ref{fig:f-idm} shows contour plots of the sensitivity region at FASER 2 for 
the fermion inelastic dark matter 
with the mass spectrum 1 (left) and 2 (right) given in \eqref{eq:spectrum-f}. 
Red, blue, green and purple contours correspond to the expected number of the signal events $3$ (95\% C.L.),$~10,~10^2$ and $10^3$, respectively, and the black one to the relic abundance of the dark matter $\Omega_{\mathrm{DM}} h^2=0.1$~\cite{ParticleDataGroup:2020ssz}.
In this case we find that annihilation process $\chi_1 \chi_1 \to A' \to \bar f f$ plays dominant role in relic density calculation.
Note that region above(below) black curve correspond to $\Omega_{\mathrm{DM}} h^2 <(>) 0.1$.
The filled color regions are excluded by the invisible decay search of the dark photon by 
NA64 (red) \cite{Banerjee:2019pds} and BaBar (green) \cite{BaBar:2017tiz}, which we rescaled according 
to our sample spectra, and dashed light blue and green curves are the limit from the E137 \cite{Batell:2014mga} 
and LSND \cite{LSND:2001akn} for reference\footnote{We rescale this 
curves from Fig.~6 of \cite{Izaguirre:2017bqb}. The spectrum is different from but similar to our spectrum.}. 
Yellow dashed line is the projection of the sensitivity at Belle-II \cite{Belle-II:2018jsg}. 
Orange band is the favored region of muon anomalous magnetic moment within $2\sigma$.

In both panels, one can see that the small kinetic mixing below the BaBar exclusion region can be explored 
by the FASER experiment. The sensitivity regions at $95$\% C.L. (red curve) reach to $\epsilon \sim 
\mathcal{O}(10^{-4})$ in case 1 and $\epsilon \sim  \mathcal{O}(10^{-5})$ in case 2. 
The sensitivity region at $95$\% C.L. (red curve) covers the smaller kinetic 
mixing below the projection of Belle-II sensitivity.
For case 2, the parameter region where $\chi_1$ satisfies the relic dark matter abundance can be examined. Larger kinetic mixing is required to satisfy the observed value of the relic abundance in case 2 than in case 1. This is simply because the number of $\chi_2$ at the freeze-out time of $\chi_1$ 
is much smaller in case 2 and hence the coannihilation mechanism does not work.

\begin{figure}[t]
\centering
\includegraphics[width=0.48\textwidth]{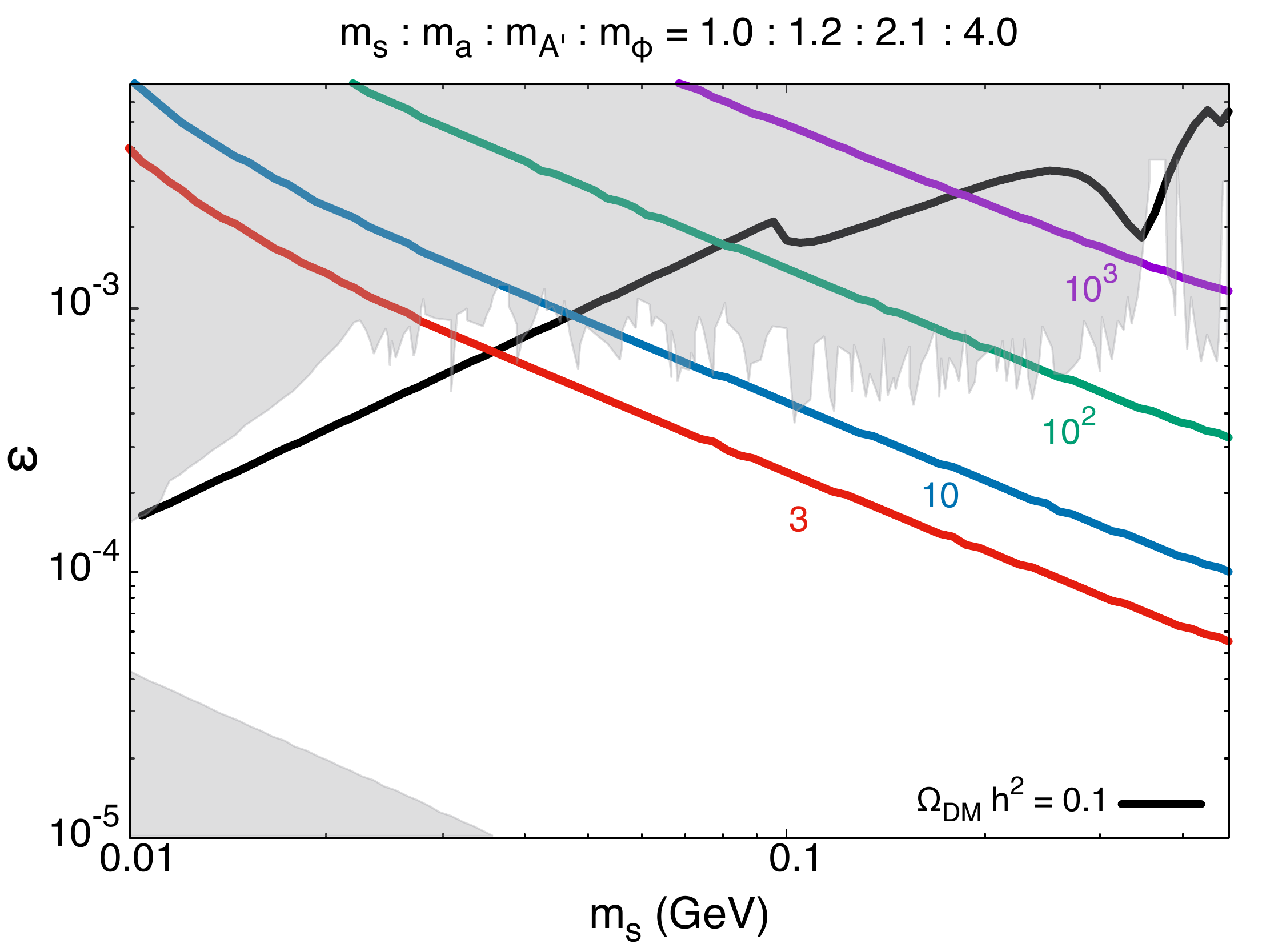} \quad
\includegraphics[width=0.48\textwidth]{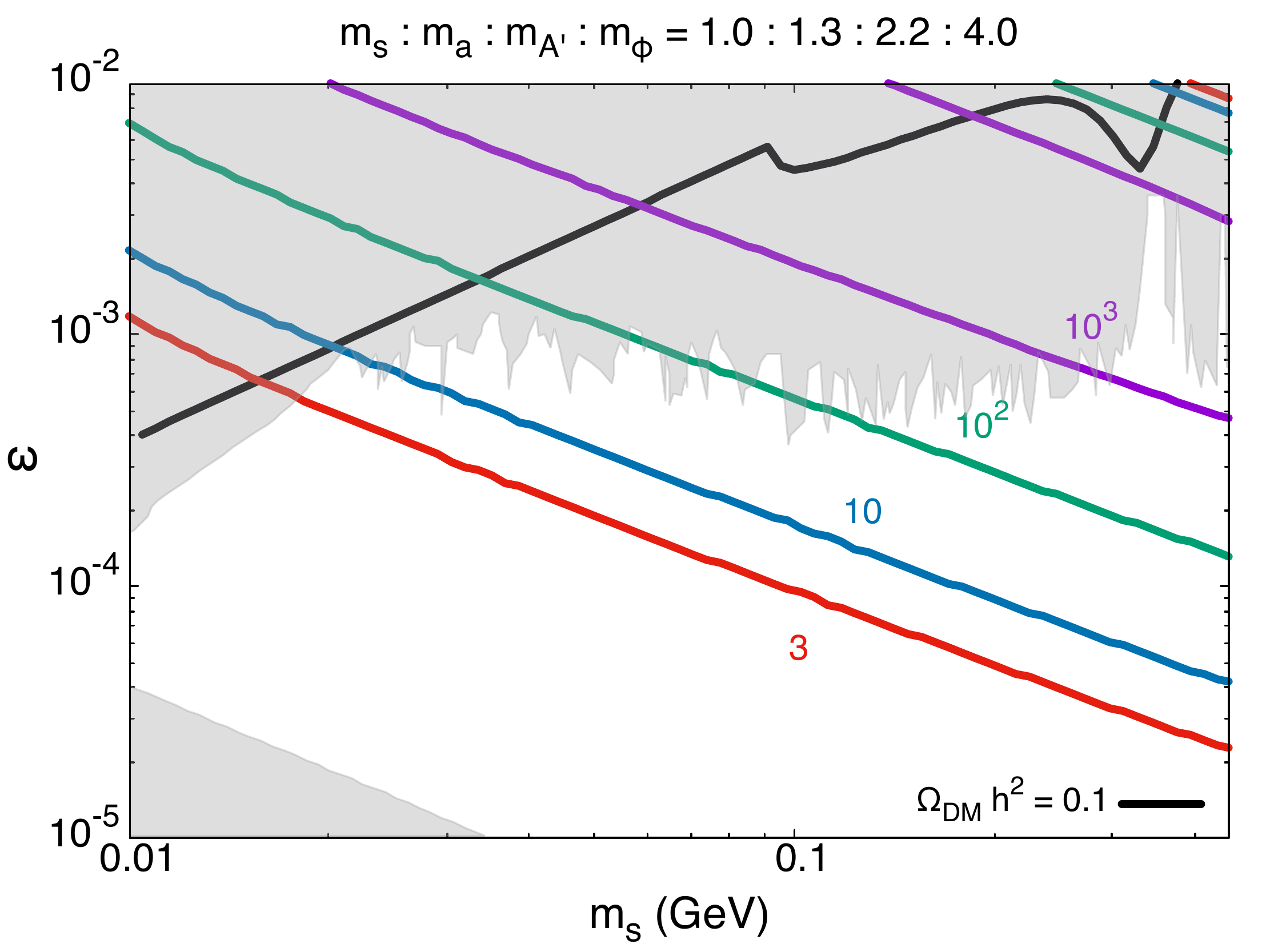}
\caption{The same figure for the scalar DM. The gray regions are taken from \cite{Jodlowski:2019ycu}.
}
\label{fig:s-idm}
\end{figure}
Figure \ref{fig:s-idm} shows the same plots for the scalar dark matter case. The gray region is exclusion region taken from \cite{Jodlowski:2019ycu}. One can see that most of the parameter region satisfying $\Omega_a h^2 < 0.1$ is already excluded or results in a few signal events. In the scalar inelastic dark matter case, 
$s$ can annihilate only through the coannihilation mechanism. With the spectrum 1 and 2, the coannihilation mechanism is less efficient, and requires to large kinetic mixing. 
The curve for $\Omega h^2 =0.1$ can be shifted to lower $\epsilon$ region if we chose mass parameters that is close to resonance $m_{A'} = m_{a} + m_{s}$ and/or smaller $m_a -m_s$. Note that the contours for number of events will shift in larger $\epsilon$ region when we make $m_a-m_s$ smaller since the lifetime 
of $\chi_2$ becomes longer.

\section{Conclusion} \label{sec:conclusion}
We have studied the inelastic dark matter scenarios in dark photon model. We incorporated a new 
scalar field which spontaneously breaks the dark gauge symmetry. Then, we considered the situation 
where the VEV of the scalar field splits the masses between the fermion or scalar dark particles. 
In such situation, the dark particles can be produced from the decay of the scalar boson 
as well as the dark photon.
 
Employing the sample spectra in which the excited dark particles are mainly produced from the decay 
of the scalar boson, we have analyzed the sensitivity of the signal events at the FASER experiment for 
the fermion and scalar inelastic dark matter. We found that the scalar boson decays provides the sizable 
number of the dark particles. We showed that the FASER 2 experiment is able to explore unconstrained 
parameter space in the fermion inelastic dark matter scenario. On the other hand, in the scalar inelastic 
dark matter scenario, we found that most of the parameter space consistent with the dark matter relic abundance is excluded by existing experiments for our choice of the mass spectra.

\clearpage
\section*{Acknowledgments}
This work is supported by JSPS KAKENHI Grant No.~18K03651,~18H01210 and 
MEXT KAKENHI Grant No.~18H05543 (T.~S.), by the Fundamental Research Funds for the Central Universities (T.~N., J.~L.), by the National Natural Science Foundation of China (NNSFC) under grant number 11905149 (J.~L.).

\bibliography{biblio}

\begin{thebibliography}{49}%
\makeatletter
\providecommand \@ifxundefined [1]{%
 \@ifx{#1\undefined}
}%
\providecommand \@ifnum [1]{%
 \ifnum #1\expandafter \@firstoftwo
 \else \expandafter \@secondoftwo
 \fi
}%
\providecommand \@ifx [1]{%
 \ifx #1\expandafter \@firstoftwo
 \else \expandafter \@secondoftwo
 \fi
}%
\providecommand \natexlab [1]{#1}%
\providecommand \enquote  [1]{``#1''}%
\providecommand \bibnamefont  [1]{#1}%
\providecommand \bibfnamefont [1]{#1}%
\providecommand \citenamefont [1]{#1}%
\providecommand \href@noop [0]{\@secondoftwo}%
\providecommand \href [0]{\begingroup \@sanitize@url \@href}%
\providecommand \@href[1]{\@@startlink{#1}\@@href}%
\providecommand \@@href[1]{\endgroup#1\@@endlink}%
\providecommand \@sanitize@url [0]{\catcode `\\12\catcode `\$12\catcode
  `\&12\catcode `\#12\catcode `\^12\catcode `\_12\catcode `\%12\relax}%
\providecommand \@@startlink[1]{}%
\providecommand \@@endlink[0]{}%
\providecommand \url  [0]{\begingroup\@sanitize@url \@url }%
\providecommand \@url [1]{\endgroup\@href {#1}{\urlprefix }}%
\providecommand \urlprefix  [0]{URL }%
\providecommand \Eprint [0]{\href }%
\providecommand \doibase [0]{http://dx.doi.org/}%
\providecommand \selectlanguage [0]{\@gobble}%
\providecommand \bibinfo  [0]{\@secondoftwo}%
\providecommand \bibfield  [0]{\@secondoftwo}%
\providecommand \translation [1]{[#1]}%
\providecommand \BibitemOpen [0]{}%
\providecommand \bibitemStop [0]{}%
\providecommand \bibitemNoStop [0]{.\EOS\space}%
\providecommand \EOS [0]{\spacefactor3000\relax}%
\providecommand \BibitemShut  [1]{\csname bibitem#1\endcsname}%
\let\auto@bib@innerbib\@empty
\bibitem [{\citenamefont {Aghanim}\ \emph {et~al.}(2020)\citenamefont {Aghanim}
  \emph {et~al.}}]{Planck:2018vyg}%
  \BibitemOpen
  \bibfield  {author} {\bibinfo {author} {\bibfnamefont {N.}~\bibnamefont
  {Aghanim}} \emph {et~al.} (\bibinfo {collaboration} {Planck}),\ }\href
  {\doibase 10.1051/0004-6361/201833910} {\bibfield  {journal} {\bibinfo
  {journal} {Astron. Astrophys.}\ }\textbf {\bibinfo {volume} {641}},\ \bibinfo
  {pages} {A6} (\bibinfo {year} {2020})},\ \bibinfo {note} {[Erratum:
  Astron.Astrophys. 652, C4 (2021)]},\ \Eprint
  {http://arxiv.org/abs/1807.06209} {arXiv:1807.06209 [astro-ph.CO]}
  \BibitemShut {NoStop}%
\bibitem [{\citenamefont {Battaglieri}\ \emph {et~al.}(2017)\citenamefont
  {Battaglieri} \emph {et~al.}}]{Battaglieri:2017aum}%
  \BibitemOpen
  \bibfield  {author} {\bibinfo {author} {\bibfnamefont {M.}~\bibnamefont
  {Battaglieri}} \emph {et~al.},\ }in\ \href@noop {} {\emph {\bibinfo
  {booktitle} {{U.S. Cosmic Visions: New Ideas in Dark Matter}}}}\ (\bibinfo
  {year} {2017})\ \Eprint {http://arxiv.org/abs/1707.04591} {arXiv:1707.04591
  [hep-ph]} \BibitemShut {NoStop}%
\bibitem [{\citenamefont {Tucker-Smith}\ and\ \citenamefont
  {Weiner}(2001)}]{Tucker-Smith:2001myb}%
  \BibitemOpen
  \bibfield  {author} {\bibinfo {author} {\bibfnamefont {D.}~\bibnamefont
  {Tucker-Smith}}\ and\ \bibinfo {author} {\bibfnamefont {N.}~\bibnamefont
  {Weiner}},\ }\href {\doibase 10.1103/PhysRevD.64.043502} {\bibfield
  {journal} {\bibinfo  {journal} {Phys. Rev. D}\ }\textbf {\bibinfo {volume}
  {64}},\ \bibinfo {pages} {043502} (\bibinfo {year} {2001})},\ \Eprint
  {http://arxiv.org/abs/hep-ph/0101138} {arXiv:hep-ph/0101138} \BibitemShut
  {NoStop}%
\bibitem [{\citenamefont {Tucker-Smith}\ and\ \citenamefont
  {Weiner}(2005)}]{Tucker-Smith:2004mxa}%
  \BibitemOpen
  \bibfield  {author} {\bibinfo {author} {\bibfnamefont {D.}~\bibnamefont
  {Tucker-Smith}}\ and\ \bibinfo {author} {\bibfnamefont {N.}~\bibnamefont
  {Weiner}},\ }\href {\doibase 10.1103/PhysRevD.72.063509} {\bibfield
  {journal} {\bibinfo  {journal} {Phys. Rev. D}\ }\textbf {\bibinfo {volume}
  {72}},\ \bibinfo {pages} {063509} (\bibinfo {year} {2005})},\ \Eprint
  {http://arxiv.org/abs/hep-ph/0402065} {arXiv:hep-ph/0402065} \BibitemShut
  {NoStop}%
\bibitem [{\citenamefont {Bernabei}\ \emph {et~al.}(2008)\citenamefont
  {Bernabei} \emph {et~al.}}]{DAMA:2008jlt}%
  \BibitemOpen
  \bibfield  {author} {\bibinfo {author} {\bibfnamefont {R.}~\bibnamefont
  {Bernabei}} \emph {et~al.} (\bibinfo {collaboration} {DAMA}),\ }\href
  {\doibase 10.1140/epjc/s10052-008-0662-y} {\bibfield  {journal} {\bibinfo
  {journal} {Eur. Phys. J. C}\ }\textbf {\bibinfo {volume} {56}},\ \bibinfo
  {pages} {333} (\bibinfo {year} {2008})},\ \Eprint
  {http://arxiv.org/abs/0804.2741} {arXiv:0804.2741 [astro-ph]} \BibitemShut
  {NoStop}%
\bibitem [{\citenamefont {Bernabei}\ \emph {et~al.}(2013)\citenamefont
  {Bernabei} \emph {et~al.}}]{Bernabei:2013xsa}%
  \BibitemOpen
  \bibfield  {author} {\bibinfo {author} {\bibfnamefont {R.}~\bibnamefont
  {Bernabei}} \emph {et~al.},\ }\href {\doibase 10.1140/epjc/s10052-013-2648-7}
  {\bibfield  {journal} {\bibinfo  {journal} {Eur. Phys. J. C}\ }\textbf
  {\bibinfo {volume} {73}},\ \bibinfo {pages} {2648} (\bibinfo {year}
  {2013})},\ \Eprint {http://arxiv.org/abs/1308.5109} {arXiv:1308.5109
  [astro-ph.GA]} \BibitemShut {NoStop}%
\bibitem [{\citenamefont {Bernabei}\ \emph {et~al.}(2010)\citenamefont
  {Bernabei} \emph {et~al.}}]{DAMA:2010gpn}%
  \BibitemOpen
  \bibfield  {author} {\bibinfo {author} {\bibfnamefont {R.}~\bibnamefont
  {Bernabei}} \emph {et~al.} (\bibinfo {collaboration} {DAMA, LIBRA}),\ }\href
  {\doibase 10.1140/epjc/s10052-010-1303-9} {\bibfield  {journal} {\bibinfo
  {journal} {Eur. Phys. J. C}\ }\textbf {\bibinfo {volume} {67}},\ \bibinfo
  {pages} {39} (\bibinfo {year} {2010})},\ \Eprint
  {http://arxiv.org/abs/1002.1028} {arXiv:1002.1028 [astro-ph.GA]} \BibitemShut
  {NoStop}%
\bibitem [{\citenamefont {Bernabei}\ \emph {et~al.}(2018)\citenamefont
  {Bernabei}, \citenamefont {Belli}, \citenamefont {Bussolotti}, \citenamefont
  {Cappella}, \citenamefont {Caracciolo}, \citenamefont {Cerulli},
  \citenamefont {Dai}, \citenamefont {D’Angelo}, \citenamefont {Di~Marco},
  \citenamefont {He}, \citenamefont {Incicchitti}, \citenamefont {Ma},
  \citenamefont {Mattei}, \citenamefont {Merlo}, \citenamefont {Montecchia},
  \citenamefont {Sheng},\ and\ \citenamefont {Ye}}]{universe4110116}%
  \BibitemOpen
  \bibfield  {author} {\bibinfo {author} {\bibfnamefont {R.}~\bibnamefont
  {Bernabei}}, \bibinfo {author} {\bibfnamefont {P.}~\bibnamefont {Belli}},
  \bibinfo {author} {\bibfnamefont {A.}~\bibnamefont {Bussolotti}}, \bibinfo
  {author} {\bibfnamefont {F.}~\bibnamefont {Cappella}}, \bibinfo {author}
  {\bibfnamefont {V.}~\bibnamefont {Caracciolo}}, \bibinfo {author}
  {\bibfnamefont {R.}~\bibnamefont {Cerulli}}, \bibinfo {author} {\bibfnamefont
  {C.-J.}\ \bibnamefont {Dai}}, \bibinfo {author} {\bibfnamefont
  {A.}~\bibnamefont {D’Angelo}}, \bibinfo {author} {\bibfnamefont
  {A.}~\bibnamefont {Di~Marco}}, \bibinfo {author} {\bibfnamefont {H.-L.}\
  \bibnamefont {He}}, \bibinfo {author} {\bibfnamefont {A.}~\bibnamefont
  {Incicchitti}}, \bibinfo {author} {\bibfnamefont {X.-H.}\ \bibnamefont {Ma}},
  \bibinfo {author} {\bibfnamefont {A.}~\bibnamefont {Mattei}}, \bibinfo
  {author} {\bibfnamefont {V.}~\bibnamefont {Merlo}}, \bibinfo {author}
  {\bibfnamefont {F.}~\bibnamefont {Montecchia}}, \bibinfo {author}
  {\bibfnamefont {X.-D.}\ \bibnamefont {Sheng}}, \ and\ \bibinfo {author}
  {\bibfnamefont {Z.-P.}\ \bibnamefont {Ye}},\ }\href {\doibase
  10.3390/universe4110116} {\bibfield  {journal} {\bibinfo  {journal}
  {Universe}\ }\textbf {\bibinfo {volume} {4}} (\bibinfo {year} {2018}),\
  10.3390/universe4110116}\BibitemShut {NoStop}%
\bibitem [{\citenamefont {Harigaya}\ \emph {et~al.}(2020)\citenamefont
  {Harigaya}, \citenamefont {Nakai},\ and\ \citenamefont
  {Suzuki}}]{Harigaya:2020ckz}%
  \BibitemOpen
  \bibfield  {author} {\bibinfo {author} {\bibfnamefont {K.}~\bibnamefont
  {Harigaya}}, \bibinfo {author} {\bibfnamefont {Y.}~\bibnamefont {Nakai}}, \
  and\ \bibinfo {author} {\bibfnamefont {M.}~\bibnamefont {Suzuki}},\ }\href
  {\doibase 10.1016/j.physletb.2020.135729} {\bibfield  {journal} {\bibinfo
  {journal} {Phys. Lett. B}\ }\textbf {\bibinfo {volume} {809}},\ \bibinfo
  {pages} {135729} (\bibinfo {year} {2020})},\ \Eprint
  {http://arxiv.org/abs/2006.11938} {arXiv:2006.11938 [hep-ph]} \BibitemShut
  {NoStop}%
\bibitem [{\citenamefont {Baek}\ \emph {et~al.}(2020)\citenamefont {Baek},
  \citenamefont {Kim},\ and\ \citenamefont {Ko}}]{Baek:2020owl}%
  \BibitemOpen
  \bibfield  {author} {\bibinfo {author} {\bibfnamefont {S.}~\bibnamefont
  {Baek}}, \bibinfo {author} {\bibfnamefont {J.}~\bibnamefont {Kim}}, \ and\
  \bibinfo {author} {\bibfnamefont {P.}~\bibnamefont {Ko}},\ }\href {\doibase
  10.1016/j.physletb.2020.135848} {\bibfield  {journal} {\bibinfo  {journal}
  {Phys. Lett. B}\ }\textbf {\bibinfo {volume} {810}},\ \bibinfo {pages}
  {135848} (\bibinfo {year} {2020})},\ \Eprint
  {http://arxiv.org/abs/2006.16876} {arXiv:2006.16876 [hep-ph]} \BibitemShut
  {NoStop}%
\bibitem [{\citenamefont {Kim}\ \emph {et~al.}(2020)\citenamefont {Kim},
  \citenamefont {Nomura},\ and\ \citenamefont {Okada}}]{Kim:2020aua}%
  \BibitemOpen
  \bibfield  {author} {\bibinfo {author} {\bibfnamefont {J.}~\bibnamefont
  {Kim}}, \bibinfo {author} {\bibfnamefont {T.}~\bibnamefont {Nomura}}, \ and\
  \bibinfo {author} {\bibfnamefont {H.}~\bibnamefont {Okada}},\ }\href
  {\doibase 10.1016/j.physletb.2020.135862} {\bibfield  {journal} {\bibinfo
  {journal} {Phys. Lett. B}\ }\textbf {\bibinfo {volume} {811}},\ \bibinfo
  {pages} {135862} (\bibinfo {year} {2020})},\ \Eprint
  {http://arxiv.org/abs/2007.09894} {arXiv:2007.09894 [hep-ph]} \BibitemShut
  {NoStop}%
\bibitem [{\citenamefont {Borah}\ \emph {et~al.}(2020)\citenamefont {Borah},
  \citenamefont {Mahapatra}, \citenamefont {Nanda},\ and\ \citenamefont
  {Sahu}}]{Borah:2020jzi}%
  \BibitemOpen
  \bibfield  {author} {\bibinfo {author} {\bibfnamefont {D.}~\bibnamefont
  {Borah}}, \bibinfo {author} {\bibfnamefont {S.}~\bibnamefont {Mahapatra}},
  \bibinfo {author} {\bibfnamefont {D.}~\bibnamefont {Nanda}}, \ and\ \bibinfo
  {author} {\bibfnamefont {N.}~\bibnamefont {Sahu}},\ }\href {\doibase
  10.1016/j.physletb.2020.135933} {\bibfield  {journal} {\bibinfo  {journal}
  {Phys. Lett. B}\ }\textbf {\bibinfo {volume} {811}},\ \bibinfo {pages}
  {135933} (\bibinfo {year} {2020})},\ \Eprint
  {http://arxiv.org/abs/2007.10754} {arXiv:2007.10754 [hep-ph]} \BibitemShut
  {NoStop}%
\bibitem [{\citenamefont {Baek}(2021)}]{Baek:2021yos}%
  \BibitemOpen
  \bibfield  {author} {\bibinfo {author} {\bibfnamefont {S.}~\bibnamefont
  {Baek}},\ }\href {\doibase 10.1007/JHEP10(2021)135} {\bibfield  {journal}
  {\bibinfo  {journal} {JHEP}\ }\textbf {\bibinfo {volume} {10}},\ \bibinfo
  {pages} {135} (\bibinfo {year} {2021})},\ \Eprint
  {http://arxiv.org/abs/2105.00877} {arXiv:2105.00877 [hep-ph]} \BibitemShut
  {NoStop}%
\bibitem [{\citenamefont {Dutta}\ \emph {et~al.}(2021)\citenamefont {Dutta},
  \citenamefont {Mahapatra}, \citenamefont {Borah},\ and\ \citenamefont
  {Sahu}}]{Dutta:2021wbn}%
  \BibitemOpen
  \bibfield  {author} {\bibinfo {author} {\bibfnamefont {M.}~\bibnamefont
  {Dutta}}, \bibinfo {author} {\bibfnamefont {S.}~\bibnamefont {Mahapatra}},
  \bibinfo {author} {\bibfnamefont {D.}~\bibnamefont {Borah}}, \ and\ \bibinfo
  {author} {\bibfnamefont {N.}~\bibnamefont {Sahu}},\ }\href {\doibase
  10.1103/PhysRevD.103.095018} {\bibfield  {journal} {\bibinfo  {journal}
  {Phys. Rev. D}\ }\textbf {\bibinfo {volume} {103}},\ \bibinfo {pages}
  {095018} (\bibinfo {year} {2021})},\ \Eprint
  {http://arxiv.org/abs/2101.06472} {arXiv:2101.06472 [hep-ph]} \BibitemShut
  {NoStop}%
\bibitem [{\citenamefont {Lee}\ and\ \citenamefont
  {Weinberg}(1977)}]{Lee:1977ua}%
  \BibitemOpen
  \bibfield  {author} {\bibinfo {author} {\bibfnamefont {B.~W.}\ \bibnamefont
  {Lee}}\ and\ \bibinfo {author} {\bibfnamefont {S.}~\bibnamefont {Weinberg}},\
  }\href {\doibase 10.1103/PhysRevLett.39.165} {\bibfield  {journal} {\bibinfo
  {journal} {Phys. Rev. Lett.}\ }\textbf {\bibinfo {volume} {39}},\ \bibinfo
  {pages} {165} (\bibinfo {year} {1977})}\BibitemShut {NoStop}%
\bibitem [{\citenamefont {Holdom}(1986)}]{Holdom:1985ag}%
  \BibitemOpen
  \bibfield  {author} {\bibinfo {author} {\bibfnamefont {B.}~\bibnamefont
  {Holdom}},\ }\href {\doibase 10.1016/0370-2693(86)91377-8} {\bibfield
  {journal} {\bibinfo  {journal} {Phys. Lett. B}\ }\textbf {\bibinfo {volume}
  {166}},\ \bibinfo {pages} {196} (\bibinfo {year} {1986})}\BibitemShut
  {NoStop}%
\bibitem [{\citenamefont {Fayet}(1990)}]{Fayet:1990wx}%
  \BibitemOpen
  \bibfield  {author} {\bibinfo {author} {\bibfnamefont {P.}~\bibnamefont
  {Fayet}},\ }\href {\doibase 10.1016/0550-3213(90)90381-M} {\bibfield
  {journal} {\bibinfo  {journal} {Nucl. Phys. B}\ }\textbf {\bibinfo {volume}
  {347}},\ \bibinfo {pages} {743} (\bibinfo {year} {1990})}\BibitemShut
  {NoStop}%
\bibitem [{\citenamefont {Bauer}\ \emph {et~al.}(2018)\citenamefont {Bauer},
  \citenamefont {Foldenauer},\ and\ \citenamefont {Jaeckel}}]{Bauer:2018onh}%
  \BibitemOpen
  \bibfield  {author} {\bibinfo {author} {\bibfnamefont {M.}~\bibnamefont
  {Bauer}}, \bibinfo {author} {\bibfnamefont {P.}~\bibnamefont {Foldenauer}}, \
  and\ \bibinfo {author} {\bibfnamefont {J.}~\bibnamefont {Jaeckel}},\ }\href
  {\doibase 10.1007/JHEP07(2018)094} {\bibfield  {journal} {\bibinfo  {journal}
  {JHEP}\ }\textbf {\bibinfo {volume} {07}},\ \bibinfo {pages} {094} (\bibinfo
  {year} {2018})},\ \Eprint {http://arxiv.org/abs/1803.05466} {arXiv:1803.05466
  [hep-ph]} \BibitemShut {NoStop}%
\bibitem [{\citenamefont {Bjorken}\ \emph {et~al.}(1988)\citenamefont
  {Bjorken}, \citenamefont {Ecklund}, \citenamefont {Nelson}, \citenamefont
  {Abashian}, \citenamefont {Church}, \citenamefont {Lu}, \citenamefont {Mo},
  \citenamefont {Nunamaker},\ and\ \citenamefont
  {Rassmann}}]{PhysRevD.38.3375}%
  \BibitemOpen
  \bibfield  {author} {\bibinfo {author} {\bibfnamefont {J.~D.}\ \bibnamefont
  {Bjorken}}, \bibinfo {author} {\bibfnamefont {S.}~\bibnamefont {Ecklund}},
  \bibinfo {author} {\bibfnamefont {W.~R.}\ \bibnamefont {Nelson}}, \bibinfo
  {author} {\bibfnamefont {A.}~\bibnamefont {Abashian}}, \bibinfo {author}
  {\bibfnamefont {C.}~\bibnamefont {Church}}, \bibinfo {author} {\bibfnamefont
  {B.}~\bibnamefont {Lu}}, \bibinfo {author} {\bibfnamefont {L.~W.}\
  \bibnamefont {Mo}}, \bibinfo {author} {\bibfnamefont {T.~A.}\ \bibnamefont
  {Nunamaker}}, \ and\ \bibinfo {author} {\bibfnamefont {P.}~\bibnamefont
  {Rassmann}},\ }\href {\doibase 10.1103/PhysRevD.38.3375} {\bibfield
  {journal} {\bibinfo  {journal} {Phys. Rev. D}\ }\textbf {\bibinfo {volume}
  {38}},\ \bibinfo {pages} {3375} (\bibinfo {year} {1988})}\BibitemShut
  {NoStop}%
\bibitem [{\citenamefont {Darm\'e}\ \emph {et~al.}(2018)\citenamefont
  {Darm\'e}, \citenamefont {Rao},\ and\ \citenamefont
  {Roszkowski}}]{Darme:2018jmx}%
  \BibitemOpen
  \bibfield  {author} {\bibinfo {author} {\bibfnamefont {L.}~\bibnamefont
  {Darm\'e}}, \bibinfo {author} {\bibfnamefont {S.}~\bibnamefont {Rao}}, \ and\
  \bibinfo {author} {\bibfnamefont {L.}~\bibnamefont {Roszkowski}},\ }\href
  {\doibase 10.1007/JHEP12(2018)014} {\bibfield  {journal} {\bibinfo  {journal}
  {JHEP}\ }\textbf {\bibinfo {volume} {12}},\ \bibinfo {pages} {014} (\bibinfo
  {year} {2018})},\ \Eprint {http://arxiv.org/abs/1807.10314} {arXiv:1807.10314
  [hep-ph]} \BibitemShut {NoStop}%
\bibitem [{\citenamefont {Auerbach}\ \emph {et~al.}(2001)\citenamefont
  {Auerbach} \emph {et~al.}}]{LSND:2001akn}%
  \BibitemOpen
  \bibfield  {author} {\bibinfo {author} {\bibfnamefont {L.~B.}\ \bibnamefont
  {Auerbach}} \emph {et~al.} (\bibinfo {collaboration} {LSND}),\ }\href
  {\doibase 10.1103/PhysRevD.63.112001} {\bibfield  {journal} {\bibinfo
  {journal} {Phys. Rev. D}\ }\textbf {\bibinfo {volume} {63}},\ \bibinfo
  {pages} {112001} (\bibinfo {year} {2001})},\ \Eprint
  {http://arxiv.org/abs/hep-ex/0101039} {arXiv:hep-ex/0101039} \BibitemShut
  {NoStop}%
\bibitem [{\citenamefont {Lees}\ \emph {et~al.}(2017)\citenamefont {Lees} \emph
  {et~al.}}]{BaBar:2017tiz}%
  \BibitemOpen
  \bibfield  {author} {\bibinfo {author} {\bibfnamefont {J.~P.}\ \bibnamefont
  {Lees}} \emph {et~al.} (\bibinfo {collaboration} {BaBar}),\ }\href {\doibase
  10.1103/PhysRevLett.119.131804} {\bibfield  {journal} {\bibinfo  {journal}
  {Phys. Rev. Lett.}\ }\textbf {\bibinfo {volume} {119}},\ \bibinfo {pages}
  {131804} (\bibinfo {year} {2017})},\ \Eprint
  {http://arxiv.org/abs/1702.03327} {arXiv:1702.03327 [hep-ex]} \BibitemShut
  {NoStop}%
\bibitem [{\citenamefont {Banerjee}\ \emph {et~al.}(2018)\citenamefont
  {Banerjee} \emph {et~al.}}]{NA64:2017vtt}%
  \BibitemOpen
  \bibfield  {author} {\bibinfo {author} {\bibfnamefont {D.}~\bibnamefont
  {Banerjee}} \emph {et~al.} (\bibinfo {collaboration} {NA64}),\ }\href
  {\doibase 10.1103/PhysRevD.97.072002} {\bibfield  {journal} {\bibinfo
  {journal} {Phys. Rev. D}\ }\textbf {\bibinfo {volume} {97}},\ \bibinfo
  {pages} {072002} (\bibinfo {year} {2018})},\ \Eprint
  {http://arxiv.org/abs/1710.00971} {arXiv:1710.00971 [hep-ex]} \BibitemShut
  {NoStop}%
\bibitem [{\citenamefont {Banerjee}\ \emph {et~al.}(2019)\citenamefont
  {Banerjee} \emph {et~al.}}]{Banerjee:2019pds}%
  \BibitemOpen
  \bibfield  {author} {\bibinfo {author} {\bibfnamefont {D.}~\bibnamefont
  {Banerjee}} \emph {et~al.},\ }\href {\doibase 10.1103/PhysRevLett.123.121801}
  {\bibfield  {journal} {\bibinfo  {journal} {Phys. Rev. Lett.}\ }\textbf
  {\bibinfo {volume} {123}},\ \bibinfo {pages} {121801} (\bibinfo {year}
  {2019})},\ \Eprint {http://arxiv.org/abs/1906.00176} {arXiv:1906.00176
  [hep-ex]} \BibitemShut {NoStop}%
\bibitem [{\citenamefont {Batell}\ \emph {et~al.}(2014)\citenamefont {Batell},
  \citenamefont {Essig},\ and\ \citenamefont {Surujon}}]{Batell:2014mga}%
  \BibitemOpen
  \bibfield  {author} {\bibinfo {author} {\bibfnamefont {B.}~\bibnamefont
  {Batell}}, \bibinfo {author} {\bibfnamefont {R.}~\bibnamefont {Essig}}, \
  and\ \bibinfo {author} {\bibfnamefont {Z.}~\bibnamefont {Surujon}},\ }\href
  {\doibase 10.1103/PhysRevLett.113.171802} {\bibfield  {journal} {\bibinfo
  {journal} {Phys. Rev. Lett.}\ }\textbf {\bibinfo {volume} {113}},\ \bibinfo
  {pages} {171802} (\bibinfo {year} {2014})},\ \Eprint
  {http://arxiv.org/abs/1406.2698} {arXiv:1406.2698 [hep-ph]} \BibitemShut
  {NoStop}%
\bibitem [{\citenamefont {Jod\l{}owski}\ \emph {et~al.}(2020)\citenamefont
  {Jod\l{}owski}, \citenamefont {Kling}, \citenamefont {Roszkowski},\ and\
  \citenamefont {Trojanowski}}]{Jodlowski:2019ycu}%
  \BibitemOpen
  \bibfield  {author} {\bibinfo {author} {\bibfnamefont {K.}~\bibnamefont
  {Jod\l{}owski}}, \bibinfo {author} {\bibfnamefont {F.}~\bibnamefont {Kling}},
  \bibinfo {author} {\bibfnamefont {L.}~\bibnamefont {Roszkowski}}, \ and\
  \bibinfo {author} {\bibfnamefont {S.}~\bibnamefont {Trojanowski}},\ }\href
  {\doibase 10.1103/PhysRevD.101.095020} {\bibfield  {journal} {\bibinfo
  {journal} {Phys. Rev. D}\ }\textbf {\bibinfo {volume} {101}},\ \bibinfo
  {pages} {095020} (\bibinfo {year} {2020})},\ \Eprint
  {http://arxiv.org/abs/1911.11346} {arXiv:1911.11346 [hep-ph]} \BibitemShut
  {NoStop}%
\bibitem [{\citenamefont {Izaguirre}\ \emph {et~al.}(2016)\citenamefont
  {Izaguirre}, \citenamefont {Krnjaic},\ and\ \citenamefont
  {Shuve}}]{Izaguirre:2015zva}%
  \BibitemOpen
  \bibfield  {author} {\bibinfo {author} {\bibfnamefont {E.}~\bibnamefont
  {Izaguirre}}, \bibinfo {author} {\bibfnamefont {G.}~\bibnamefont {Krnjaic}},
  \ and\ \bibinfo {author} {\bibfnamefont {B.}~\bibnamefont {Shuve}},\ }\href
  {\doibase 10.1103/PhysRevD.93.063523} {\bibfield  {journal} {\bibinfo
  {journal} {Phys. Rev. D}\ }\textbf {\bibinfo {volume} {93}},\ \bibinfo
  {pages} {063523} (\bibinfo {year} {2016})},\ \Eprint
  {http://arxiv.org/abs/1508.03050} {arXiv:1508.03050 [hep-ph]} \BibitemShut
  {NoStop}%
\bibitem [{\citenamefont {Duerr}\ \emph {et~al.}(2020)\citenamefont {Duerr},
  \citenamefont {Ferber}, \citenamefont {Hearty}, \citenamefont {Kahlhoefer},
  \citenamefont {Schmidt-Hoberg},\ and\ \citenamefont
  {Tunney}}]{Duerr:2019dmv}%
  \BibitemOpen
  \bibfield  {author} {\bibinfo {author} {\bibfnamefont {M.}~\bibnamefont
  {Duerr}}, \bibinfo {author} {\bibfnamefont {T.}~\bibnamefont {Ferber}},
  \bibinfo {author} {\bibfnamefont {C.}~\bibnamefont {Hearty}}, \bibinfo
  {author} {\bibfnamefont {F.}~\bibnamefont {Kahlhoefer}}, \bibinfo {author}
  {\bibfnamefont {K.}~\bibnamefont {Schmidt-Hoberg}}, \ and\ \bibinfo {author}
  {\bibfnamefont {P.}~\bibnamefont {Tunney}},\ }\href {\doibase
  10.1007/JHEP02(2020)039} {\bibfield  {journal} {\bibinfo  {journal} {JHEP}\
  }\textbf {\bibinfo {volume} {02}},\ \bibinfo {pages} {039} (\bibinfo {year}
  {2020})},\ \Eprint {http://arxiv.org/abs/1911.03176} {arXiv:1911.03176
  [hep-ph]} \BibitemShut {NoStop}%
\bibitem [{\citenamefont {Duerr}\ \emph {et~al.}(2021)\citenamefont {Duerr},
  \citenamefont {Ferber}, \citenamefont {Garcia-Cely}, \citenamefont {Hearty},\
  and\ \citenamefont {Schmidt-Hoberg}}]{Duerr:2020muu}%
  \BibitemOpen
  \bibfield  {author} {\bibinfo {author} {\bibfnamefont {M.}~\bibnamefont
  {Duerr}}, \bibinfo {author} {\bibfnamefont {T.}~\bibnamefont {Ferber}},
  \bibinfo {author} {\bibfnamefont {C.}~\bibnamefont {Garcia-Cely}}, \bibinfo
  {author} {\bibfnamefont {C.}~\bibnamefont {Hearty}}, \ and\ \bibinfo {author}
  {\bibfnamefont {K.}~\bibnamefont {Schmidt-Hoberg}},\ }\href {\doibase
  10.1007/JHEP04(2021)146} {\bibfield  {journal} {\bibinfo  {journal} {JHEP}\
  }\textbf {\bibinfo {volume} {04}},\ \bibinfo {pages} {146} (\bibinfo {year}
  {2021})},\ \Eprint {http://arxiv.org/abs/2012.08595} {arXiv:2012.08595
  [hep-ph]} \BibitemShut {NoStop}%
\bibitem [{\citenamefont {Kang}\ \emph {et~al.}(2021)\citenamefont {Kang},
  \citenamefont {Ko},\ and\ \citenamefont {Lu}}]{Kang:2021oes}%
  \BibitemOpen
  \bibfield  {author} {\bibinfo {author} {\bibfnamefont {D.~W.}\ \bibnamefont
  {Kang}}, \bibinfo {author} {\bibfnamefont {P.}~\bibnamefont {Ko}}, \ and\
  \bibinfo {author} {\bibfnamefont {C.-T.}\ \bibnamefont {Lu}},\ }\href
  {\doibase 10.1007/JHEP04(2021)269} {\bibfield  {journal} {\bibinfo  {journal}
  {JHEP}\ }\textbf {\bibinfo {volume} {04}},\ \bibinfo {pages} {269} (\bibinfo
  {year} {2021})},\ \Eprint {http://arxiv.org/abs/2101.02503} {arXiv:2101.02503
  [hep-ph]} \BibitemShut {NoStop}%
\bibitem [{\citenamefont {Dreyer}\ \emph {et~al.}(2021)\citenamefont {Dreyer}
  \emph {et~al.}}]{Dreyer:2021aqd}%
  \BibitemOpen
  \bibfield  {author} {\bibinfo {author} {\bibfnamefont {S.}~\bibnamefont
  {Dreyer}} \emph {et~al.},\ }\href@noop {} {\  (\bibinfo {year} {2021})},\
  \Eprint {http://arxiv.org/abs/2105.12962} {arXiv:2105.12962 [hep-ph]}
  \BibitemShut {NoStop}%
\bibitem [{\citenamefont {Berlin}\ and\ \citenamefont
  {Kling}(2019)}]{Berlin:2018jbm}%
  \BibitemOpen
  \bibfield  {author} {\bibinfo {author} {\bibfnamefont {A.}~\bibnamefont
  {Berlin}}\ and\ \bibinfo {author} {\bibfnamefont {F.}~\bibnamefont {Kling}},\
  }\href {\doibase 10.1103/PhysRevD.99.015021} {\bibfield  {journal} {\bibinfo
  {journal} {Phys. Rev. D}\ }\textbf {\bibinfo {volume} {99}},\ \bibinfo
  {pages} {015021} (\bibinfo {year} {2019})},\ \Eprint
  {http://arxiv.org/abs/1810.01879} {arXiv:1810.01879 [hep-ph]} \BibitemShut
  {NoStop}%
\bibitem [{\citenamefont {Batell}\ \emph
  {et~al.}(2021{\natexlab{a}})\citenamefont {Batell}, \citenamefont {Feng},\
  and\ \citenamefont {Trojanowski}}]{Batell:2021blf}%
  \BibitemOpen
  \bibfield  {author} {\bibinfo {author} {\bibfnamefont {B.}~\bibnamefont
  {Batell}}, \bibinfo {author} {\bibfnamefont {J.~L.}\ \bibnamefont {Feng}}, \
  and\ \bibinfo {author} {\bibfnamefont {S.}~\bibnamefont {Trojanowski}},\
  }\href {\doibase 10.1103/PhysRevD.103.075023} {\bibfield  {journal} {\bibinfo
   {journal} {Phys. Rev. D}\ }\textbf {\bibinfo {volume} {103}},\ \bibinfo
  {pages} {075023} (\bibinfo {year} {2021}{\natexlab{a}})},\ \Eprint
  {http://arxiv.org/abs/2101.10338} {arXiv:2101.10338 [hep-ph]} \BibitemShut
  {NoStop}%
\bibitem [{\citenamefont {Guo}\ \emph {et~al.}(2021)\citenamefont {Guo},
  \citenamefont {He}, \citenamefont {Liu},\ and\ \citenamefont
  {Wang}}]{Guo:2021vpb}%
  \BibitemOpen
  \bibfield  {author} {\bibinfo {author} {\bibfnamefont {J.}~\bibnamefont
  {Guo}}, \bibinfo {author} {\bibfnamefont {Y.}~\bibnamefont {He}}, \bibinfo
  {author} {\bibfnamefont {J.}~\bibnamefont {Liu}}, \ and\ \bibinfo {author}
  {\bibfnamefont {X.-P.}\ \bibnamefont {Wang}},\ }\href@noop {} {\  (\bibinfo
  {year} {2021})},\ \Eprint {http://arxiv.org/abs/2111.01164} {arXiv:2111.01164
  [hep-ph]} \BibitemShut {NoStop}%
\bibitem [{\citenamefont {Batell}\ \emph
  {et~al.}(2021{\natexlab{b}})\citenamefont {Batell}, \citenamefont {Berger},
  \citenamefont {Darm\'e},\ and\ \citenamefont {Frugiuele}}]{Batell:2021ooj}%
  \BibitemOpen
  \bibfield  {author} {\bibinfo {author} {\bibfnamefont {B.}~\bibnamefont
  {Batell}}, \bibinfo {author} {\bibfnamefont {J.}~\bibnamefont {Berger}},
  \bibinfo {author} {\bibfnamefont {L.}~\bibnamefont {Darm\'e}}, \ and\
  \bibinfo {author} {\bibfnamefont {C.}~\bibnamefont {Frugiuele}},\ }\href
  {\doibase 10.1103/PhysRevD.104.075026} {\bibfield  {journal} {\bibinfo
  {journal} {Phys. Rev. D}\ }\textbf {\bibinfo {volume} {104}},\ \bibinfo
  {pages} {075026} (\bibinfo {year} {2021}{\natexlab{b}})},\ \Eprint
  {http://arxiv.org/abs/2106.04584} {arXiv:2106.04584 [hep-ph]} \BibitemShut
  {NoStop}%
\bibitem [{\citenamefont {Feng}\ \emph
  {et~al.}(2018{\natexlab{a}})\citenamefont {Feng}, \citenamefont {Galon},
  \citenamefont {Kling},\ and\ \citenamefont {Trojanowski}}]{Feng:2017uoz}%
  \BibitemOpen
  \bibfield  {author} {\bibinfo {author} {\bibfnamefont {J.~L.}\ \bibnamefont
  {Feng}}, \bibinfo {author} {\bibfnamefont {I.}~\bibnamefont {Galon}},
  \bibinfo {author} {\bibfnamefont {F.}~\bibnamefont {Kling}}, \ and\ \bibinfo
  {author} {\bibfnamefont {S.}~\bibnamefont {Trojanowski}},\ }\href {\doibase
  10.1103/PhysRevD.97.035001} {\bibfield  {journal} {\bibinfo  {journal} {Phys.
  Rev. D}\ }\textbf {\bibinfo {volume} {97}},\ \bibinfo {pages} {035001}
  (\bibinfo {year} {2018}{\natexlab{a}})},\ \Eprint
  {http://arxiv.org/abs/1708.09389} {arXiv:1708.09389 [hep-ph]} \BibitemShut
  {NoStop}%
\bibitem [{\citenamefont {Ariga}\ \emph {et~al.}(2018)\citenamefont {Ariga}
  \emph {et~al.}}]{FASER:2018bac}%
  \BibitemOpen
  \bibfield  {author} {\bibinfo {author} {\bibfnamefont {A.}~\bibnamefont
  {Ariga}} \emph {et~al.} (\bibinfo {collaboration} {FASER}),\ }\href@noop {}
  {\  (\bibinfo {year} {2018})},\ \Eprint {http://arxiv.org/abs/1812.09139}
  {arXiv:1812.09139 [physics.ins-det]} \BibitemShut {NoStop}%
\bibitem [{\citenamefont {B\'elanger}\ \emph {et~al.}(2015)\citenamefont
  {B\'elanger}, \citenamefont {Boudjema}, \citenamefont {Pukhov},\ and\
  \citenamefont {Semenov}}]{Belanger:2014vza}%
  \BibitemOpen
  \bibfield  {author} {\bibinfo {author} {\bibfnamefont {G.}~\bibnamefont
  {B\'elanger}}, \bibinfo {author} {\bibfnamefont {F.}~\bibnamefont
  {Boudjema}}, \bibinfo {author} {\bibfnamefont {A.}~\bibnamefont {Pukhov}}, \
  and\ \bibinfo {author} {\bibfnamefont {A.}~\bibnamefont {Semenov}},\ }\href
  {\doibase 10.1016/j.cpc.2015.03.003} {\bibfield  {journal} {\bibinfo
  {journal} {Comput. Phys. Commun.}\ }\textbf {\bibinfo {volume} {192}},\
  \bibinfo {pages} {322} (\bibinfo {year} {2015})},\ \Eprint
  {http://arxiv.org/abs/1407.6129} {arXiv:1407.6129 [hep-ph]} \BibitemShut
  {NoStop}%
\bibitem [{\citenamefont {Feng}\ \emph
  {et~al.}(2018{\natexlab{b}})\citenamefont {Feng}, \citenamefont {Galon},
  \citenamefont {Kling},\ and\ \citenamefont {Trojanowski}}]{Feng:2017vli}%
  \BibitemOpen
  \bibfield  {author} {\bibinfo {author} {\bibfnamefont {J.~L.}\ \bibnamefont
  {Feng}}, \bibinfo {author} {\bibfnamefont {I.}~\bibnamefont {Galon}},
  \bibinfo {author} {\bibfnamefont {F.}~\bibnamefont {Kling}}, \ and\ \bibinfo
  {author} {\bibfnamefont {S.}~\bibnamefont {Trojanowski}},\ }\href {\doibase
  10.1103/PhysRevD.97.055034} {\bibfield  {journal} {\bibinfo  {journal} {Phys.
  Rev. D}\ }\textbf {\bibinfo {volume} {97}},\ \bibinfo {pages} {055034}
  (\bibinfo {year} {2018}{\natexlab{b}})},\ \Eprint
  {http://arxiv.org/abs/1710.09387} {arXiv:1710.09387 [hep-ph]} \BibitemShut
  {NoStop}%
\bibitem [{\citenamefont {Pierog}\ \emph {et~al.}(2015)\citenamefont {Pierog},
  \citenamefont {Karpenko}, \citenamefont {Katzy}, \citenamefont {Yatsenko},\
  and\ \citenamefont {Werner}}]{Pierog:2013ria}%
  \BibitemOpen
  \bibfield  {author} {\bibinfo {author} {\bibfnamefont {T.}~\bibnamefont
  {Pierog}}, \bibinfo {author} {\bibfnamefont {I.}~\bibnamefont {Karpenko}},
  \bibinfo {author} {\bibfnamefont {J.~M.}\ \bibnamefont {Katzy}}, \bibinfo
  {author} {\bibfnamefont {E.}~\bibnamefont {Yatsenko}}, \ and\ \bibinfo
  {author} {\bibfnamefont {K.}~\bibnamefont {Werner}},\ }\href {\doibase
  10.1103/PhysRevC.92.034906} {\bibfield  {journal} {\bibinfo  {journal} {Phys.
  Rev. C}\ }\textbf {\bibinfo {volume} {92}},\ \bibinfo {pages} {034906}
  (\bibinfo {year} {2015})},\ \Eprint {http://arxiv.org/abs/1306.0121}
  {arXiv:1306.0121 [hep-ph]} \BibitemShut {NoStop}%
\bibitem [{\citenamefont {Baus}\ \emph {et~al.}(2021)\citenamefont {Baus},
  \citenamefont {Pierog},\ and\ \citenamefont {Ulrich}}]{crmc:html}%
  \BibitemOpen
  \bibfield  {author} {\bibinfo {author} {\bibfnamefont {C.}~\bibnamefont
  {Baus}}, \bibinfo {author} {\bibfnamefont {T.}~\bibnamefont {Pierog}}, \ and\
  \bibinfo {author} {\bibfnamefont {R.}~\bibnamefont {Ulrich}},\ }\href@noop {}
  {\  (\bibinfo {year} {2021})},\ \Eprint
  {http://arxiv.org/abs/https://web.ikp.kit.edu/rulrich/crmc.html}
  {https://web.ikp.kit.edu/rulrich/crmc.html} \BibitemShut {NoStop}%
\bibitem [{\citenamefont {Sjostrand}\ \emph {et~al.}(2008)\citenamefont
  {Sjostrand}, \citenamefont {Mrenna},\ and\ \citenamefont
  {Skands}}]{Sjostrand:2007gs}%
  \BibitemOpen
  \bibfield  {author} {\bibinfo {author} {\bibfnamefont {T.}~\bibnamefont
  {Sjostrand}}, \bibinfo {author} {\bibfnamefont {S.}~\bibnamefont {Mrenna}}, \
  and\ \bibinfo {author} {\bibfnamefont {P.~Z.}\ \bibnamefont {Skands}},\
  }\href {\doibase 10.1016/j.cpc.2008.01.036} {\bibfield  {journal} {\bibinfo
  {journal} {Comput. Phys. Commun.}\ }\textbf {\bibinfo {volume} {178}},\
  \bibinfo {pages} {852} (\bibinfo {year} {2008})},\ \Eprint
  {http://arxiv.org/abs/0710.3820} {arXiv:0710.3820 [hep-ph]} \BibitemShut
  {NoStop}%
\bibitem [{\citenamefont {Skands}\ \emph {et~al.}(2014)\citenamefont {Skands},
  \citenamefont {Carrazza},\ and\ \citenamefont {Rojo}}]{Skands:2014pea}%
  \BibitemOpen
  \bibfield  {author} {\bibinfo {author} {\bibfnamefont {P.}~\bibnamefont
  {Skands}}, \bibinfo {author} {\bibfnamefont {S.}~\bibnamefont {Carrazza}}, \
  and\ \bibinfo {author} {\bibfnamefont {J.}~\bibnamefont {Rojo}},\ }\href
  {\doibase 10.1140/epjc/s10052-014-3024-y} {\bibfield  {journal} {\bibinfo
  {journal} {Eur. Phys. J. C}\ }\textbf {\bibinfo {volume} {74}},\ \bibinfo
  {pages} {3024} (\bibinfo {year} {2014})},\ \Eprint
  {http://arxiv.org/abs/1404.5630} {arXiv:1404.5630 [hep-ph]} \BibitemShut
  {NoStop}%
\bibitem [{\citenamefont {ATLAS}(2016)}]{ATLAS:2016puo}%
  \BibitemOpen
  \bibfield  {author} {\bibinfo {author} {\bibnamefont {ATLAS}},\ }\href@noop
  {} {\  (\bibinfo {year} {2016})},\ \Eprint
  {http://arxiv.org/abs/ATL-PHYS-PUB-2016-017} {ATL-PHYS-PUB-2016-017}
  \BibitemShut {NoStop}%
\bibitem [{\citenamefont {Kling}\ and\ \citenamefont
  {Trojanowski}(2021)}]{Kling:2021fwx}%
  \BibitemOpen
  \bibfield  {author} {\bibinfo {author} {\bibfnamefont {F.}~\bibnamefont
  {Kling}}\ and\ \bibinfo {author} {\bibfnamefont {S.}~\bibnamefont
  {Trojanowski}},\ }\href {\doibase 10.1103/PhysRevD.104.035012} {\bibfield
  {journal} {\bibinfo  {journal} {Phys. Rev. D}\ }\textbf {\bibinfo {volume}
  {104}},\ \bibinfo {pages} {035012} (\bibinfo {year} {2021})},\ \Eprint
  {http://arxiv.org/abs/2105.07077} {arXiv:2105.07077 [hep-ph]} \BibitemShut
  {NoStop}%
\bibitem [{\citenamefont {Araki}\ \emph {et~al.}(2021)\citenamefont {Araki},
  \citenamefont {Asai}, \citenamefont {Otono}, \citenamefont {Shimomura},\ and\
  \citenamefont {Takubo}}]{Araki:2020wkq}%
  \BibitemOpen
  \bibfield  {author} {\bibinfo {author} {\bibfnamefont {T.}~\bibnamefont
  {Araki}}, \bibinfo {author} {\bibfnamefont {K.}~\bibnamefont {Asai}},
  \bibinfo {author} {\bibfnamefont {H.}~\bibnamefont {Otono}}, \bibinfo
  {author} {\bibfnamefont {T.}~\bibnamefont {Shimomura}}, \ and\ \bibinfo
  {author} {\bibfnamefont {Y.}~\bibnamefont {Takubo}},\ }\href {\doibase
  10.1007/JHEP03(2021)072} {\bibfield  {journal} {\bibinfo  {journal} {JHEP}\
  }\textbf {\bibinfo {volume} {03}},\ \bibinfo {pages} {072} (\bibinfo {year}
  {2021})},\ \bibinfo {note} {[Erratum: JHEP 06, 087 (2021)]},\ \Eprint
  {http://arxiv.org/abs/2008.12765} {arXiv:2008.12765 [hep-ph]} \BibitemShut
  {NoStop}%
\bibitem [{\citenamefont {Altmannshofer}\ \emph {et~al.}(2019)\citenamefont
  {Altmannshofer} \emph {et~al.}}]{Belle-II:2018jsg}%
  \BibitemOpen
  \bibfield  {author} {\bibinfo {author} {\bibfnamefont {W.}~\bibnamefont
  {Altmannshofer}} \emph {et~al.} (\bibinfo {collaboration} {Belle-II}),\
  }\href {\doibase 10.1093/ptep/ptz106} {\bibfield  {journal} {\bibinfo
  {journal} {PTEP}\ }\textbf {\bibinfo {volume} {2019}},\ \bibinfo {pages}
  {123C01} (\bibinfo {year} {2019})},\ \bibinfo {note} {[Erratum: PTEP 2020,
  029201 (2020)]},\ \Eprint {http://arxiv.org/abs/1808.10567} {arXiv:1808.10567
  [hep-ex]} \BibitemShut {NoStop}%
\bibitem [{\citenamefont {Zyla}\ \emph {et~al.}(2020)\citenamefont {Zyla} \emph
  {et~al.}}]{ParticleDataGroup:2020ssz}%
  \BibitemOpen
  \bibfield  {author} {\bibinfo {author} {\bibfnamefont {P.~A.}\ \bibnamefont
  {Zyla}} \emph {et~al.} (\bibinfo {collaboration} {Particle Data Group}),\
  }\href {\doibase 10.1093/ptep/ptaa104} {\bibfield  {journal} {\bibinfo
  {journal} {PTEP}\ }\textbf {\bibinfo {volume} {2020}},\ \bibinfo {pages}
  {083C01} (\bibinfo {year} {2020})}\BibitemShut {NoStop}%
\bibitem [{\citenamefont {Izaguirre}\ \emph {et~al.}(2017)\citenamefont
  {Izaguirre}, \citenamefont {Kahn}, \citenamefont {Krnjaic},\ and\
  \citenamefont {Moschella}}]{Izaguirre:2017bqb}%
  \BibitemOpen
  \bibfield  {author} {\bibinfo {author} {\bibfnamefont {E.}~\bibnamefont
  {Izaguirre}}, \bibinfo {author} {\bibfnamefont {Y.}~\bibnamefont {Kahn}},
  \bibinfo {author} {\bibfnamefont {G.}~\bibnamefont {Krnjaic}}, \ and\
  \bibinfo {author} {\bibfnamefont {M.}~\bibnamefont {Moschella}},\ }\href
  {\doibase 10.1103/PhysRevD.96.055007} {\bibfield  {journal} {\bibinfo
  {journal} {Phys. Rev. D}\ }\textbf {\bibinfo {volume} {96}},\ \bibinfo
  {pages} {055007} (\bibinfo {year} {2017})},\ \Eprint
  {http://arxiv.org/abs/1703.06881} {arXiv:1703.06881 [hep-ph]} \BibitemShut
  {NoStop}%
\end{thebibliography}%

\end{document}